\documentclass{aa}
\usepackage{graphicx}
\usepackage{supertabular}
\usepackage{aalongtable}
\usepackage{txfonts}
\usepackage{natbib}
\usepackage{lscape}
\begin{document}
\title{Infrared Photometry and Evolution of Mass-Losing AGB Stars.}
\subtitle{II. Luminosity and Colors of MS and S Stars}

\author
    {
      R. Guandalini\inst{1}
      \and
      M. Busso\inst{1}
     }

\offprints{R. Guandalini, guandalini@fisica.unipg.it}

\institute
    {Department of Physics, University of Perugia, and INFN, Sezione di Perugia, Via A. Pascoli 1,
      06123 Perugia, Italy\\
      \email{guandalini@fisica.unipg.it;busso@fisica.unipg.it}
    }

\date{Received / Accepted }


\abstract
{Asymptotic Giant Branch (AGB) phases mark the end of the
evolution for Low- and Intermediate-Mass Stars. Our understanding
of the mechanisms through which they eject the envelope and our
assessment of their contribution to the mass return to the
Interstellar Medium and to the chemical evolution of Galaxies are
hampered by poor knowledge of their Luminosities and mass loss
rates, both for C-rich and for O-rich sources.}
{We plan to establish criteria permitting a more quantitative
determination of luminosities (and subsequently of mass loss rates)
for the various types of AGB stars on the basis of infrared fluxes.
In this paper, in particular, we concentrate on O-rich and
s-element-rich MS, S stars and include a small sample of SC stars.}
{We reanalyze the absolute bolometric magnitudes and colors of MS,
S, SC stars on the basis of a sample of intrinsic (single) and
extrinsic (binary) long period variables. We derive bolometric
corrections as a function of near- and mid-infrared colors, adopting
as references a group of stars for which the Spectral Energy
Distribution could be reconstructed in detail over a large
wavelength range. We determine the absolute HR diagrams, and compare
luminosities and colors of S-type giants with those, previously
derived, of C-rich AGB stars. Luminosity estimates are also verified
on the basis of existing Period-Luminosity relations valid for
O-rich Miras.}
{S star bolometric luminosities are almost indistinguishable from
those of C-rich AGB stars. On the contrary, their circumstellar
envelopes are thinner and less opaque. Despite this last property
the IR wavelengths remain dominant, with the bluest stars having
their maximum emission in the H or K(short) bands. Near-to-Mid
infrared color differences are in any case smaller than for C stars.
Based on Period-Luminosity relations for O-rich Miras and on
Magnitude-color relations for the same variables we show how
approximate distances (hence intrinsic parameters) for sources of so
far unknown parallax can be inferred. We argue that most of the
sources have a rather small mass ($<$ 2 $M_{\odot}$); dredge-up
might then be not effective enough to let the C/O ratio exceed
unity.} {}


\titlerunning{The Evolutionary Status of Mass-Losing AGB Stars. II.}
\authorrunning{R. Guandalini \and M. Busso}

\maketitle


\section{Introduction \label{sect1}}

The Asymptotic Giant Branch phases (hereafter AGB) represent the
second ascent along the Red Giant Branch, occurring after the
exhaustion of core He burning for all stars between $\sim$~0.8 and
8.0\,M$_{\sun}$. In these evolutionary stages, stars are powered by
two nuclear shells, burning H and He alternatively. In particular,
in the final 1-2 Myr of the AGB, the He-burning shell remains mainly
quiescent, if not for recurrent explosive ignitions during which a
lot of C (from 20 to 25\% by mass) is produced and spread over the
whole He-rich layer, in short phases of convective mixing (the
so-called {\it thermal pulses}). Convective penetration of the
envelope follows, in repeated episodes collectively called ''the
third dredge-up'', and carries the new carbon to the surface,
together with other nucleosynthesis products, in particular
s-elements generated by efficient neutron captures \citep{busso99}.

\begin{table*}[t!]
\caption{Sample A $-$ First part. Spectral Type is from the GCVS
catalogue whenever possible, otherwise it is obtained from the SIMBAD Astronomical Database. }             
\label{table:1}      
\centering                          
\begin{tabular}{c c c c c c}        
\hline \hline
IRAS    &   Other   &   Stephenson  &   Coordinates &   Spectral Type   &   Var. Type   \\
name    &   name    &   name    &   ICRS    &       &   (GCVS)  \\
\hline \hline
01159+7220  &   \object{S Cas}  &   CSS 28  &   01 19 41.97 +72 36 39.3 &   S3,4e$-$S5,8e   &   M   \\
19126$-$0708    &   \object{W Aql}  &   CSS 1115    &   19 15 23.44 $-$07 02 49.9   &   S3,9e$-$S6,9e   &   M   \\
19354+5005  &   \object{R Cyg}  &   CSS 1150    &   19 36 49.381 +50 11 59.46   &   S2.5,9e$-$S6,9e(Tc) &   M   \\
19486+3247  &   \object{chi Cyg}    &   CSS 1165    &   19 50 33.9220 +32 54 50.610     &   S6,2e$-$S10,4e/MSe  &   M   \\
23595$-$1457    &   \object{W Cet}  &   CSS 1346    &   00 02 07.3891 $-$14 40 33.065   &   S6,3e$-$S9,2e   &   M   \\
22196$-$4612    &   \object{pi1 Gru}    &   CSS 1294    &   22 22 44.2053 $-$45 56 52.598   &   S5,7e   &   SRB \\
20026+3640  &   \object{AA Cyg} &   CSS 1188    &   20 04 27.6055 +36 49 00.465 &   S7,5$-$S7.5,6(MpTc) &   SRB \\
20120$-$4433    &   \object{RZ Sgr} &   CSS 1196    &   20 15 28.4049 $-$44 24 37.480   &   S4,4ep  &   SRB \\
03452+5301  &   \object{WX Cam} &   CSS 82  &   03 49 03.77 +53 10 59.2 &   S5,8    &   LB  \\
\hline
23070+0824  &   \object{GZ Peg} &   CSS 1322    &   23 09 31.4570 +08 40 37.778 &   M4SIII  &   SRA \\
15492+4837  &   \object{ST Her} &   CSS 903 &   15 50 46.6248 +48 28 58.856 &   M6$-$7IIIaS &   SRB \\
00192$-$2020    &   \object{T Cet}  &   CSS 8   &   00 21 46.2737 $-$20 03 28.885   &   M5$-$6SIIe  &   SRC \\
05374+3153  &   \object{NO Aur} &   CSS 149 &   05 40 42.0504 +31 55 14.187 &   M2SIab  &   LC  \\
22476+4047  &   \object{RX Lac} &   CSS 1308    &   22 49 56.8992 +41 03 04.312     &   M7.5Se  &   SRB \\
\hline
00213+3817  &   \object{R And}  &   CSS 9   &   00 24 01.9469 +38 34 37.328 &   S3,5e$-$S8,8e/M7e   &   M   \\
22521+1640  &   \object{HR Peg} &   CSS 1315    &   22 54 35.6272 +16 56 30.601     &   S5,1/M4 &   SRB \\
17553+4521  &   \object{OP Her} &   $-$ &   17 56 48.5274 +45 21 03.063 &   M5IIb$-$IIIa/S  &   SRB \\
\hline
13372$-$7136    &   \object{LY Mus} &   CSS 826 &   13 41 13.5883 $-$71 52 05.767   &   M4III   &   LB  \\
18058$-$3658    &   $-$ &   \object{CSS 1023}   &   18 09 17.1853 $-$36 57 57.614   &   M2II$-$III  &   $-$ \\
\hline
19111+2555  &   \object{S Lyr}  &   CSS 1112    &   19 13 11.79 +26 00 28.3 &   SCe &   M   \\
\hline
15194$-$5115    &   \object{II Lup} &   CSS 886 &   15 23 04.91 $-$51 25 59.0   &   C   &   M   \\
\hline \hline
\end{tabular}
\end{table*}

\begin{table*}[t!]
\caption{Sample A $-$ Second part.}             
\label{table:2}      
\centering                          
\begin{tabular}{c c c c c c c c c c c c}        
\hline      \hline
Source  &   J   &   H   &   K   &   [8.8]   &   [9.8]   &   [11.7]  &   [12.5]  &   D   &   E   &   Mid$-$IR Data Origin    &   ISO \\
name    &   [Jy]    &   [Jy]    &   [Jy]    &   [Jy]    &   [Jy]    &   [Jy]    &   [Jy]    &   [Jy]    &   [Jy]    &       &   TDT Number  \\
\hline      \hline
S Cas   &   56.6    &   168 &   244 &   232 &   316 &   304 &   281 &   235 &   185 &   ISO$-$SWS1  &   41602133    \\
W Aql   &   388 &   822 &   1113    &   969 &   1099    &   1043    &   789 &   625 &   488 &   ISO$-$SWS1  &   16402335    \\
R Cyg   &   200 &   288 &   302 &   77.3    &   87.8    &   79.3    &   69.2    &   52.8    &   29.7    &   ISO$-$SWS1  &   42201625    \\
chi Cyg &   1365    &   2823    &   3176    &   1408    &   1655    &   1579    &   1095    &   736 &   487 &   ISO$-$SWS1  &   15900437    \\
W Cet   &   74.1    &   113 &   101 &   12.7    &   12.4    &   10.5    &   10.9    &   8.3 &   4.5 &   ISO$-$SWS1  &   37802225    \\
pi1 Gru &   3080    &   $-$ &   5812    &   544 &   632 &   679 &   683 &   505 &   365 &   ISO$-$SWS1  &   34402039    \\
AA Cyg  &   238 &   389 &   375 &   42.6    &   42.5    &   42.9    &   36.8    &   27.5    &   16.9    &   ISO$-$SWS1  &   36401817    \\
RZ Sgr  &   139 &   206 &   190 &   25.0    &   26.2    &   25.7    &   24.5    &   21.8    &   21.0    &   ISO$-$SWS1  &   14100818    \\
WX Cam  &   43.5    &   79.3    &   87.6    &   8.1 &   7.9 &   8.4 &   16.6    &   9.3 &   4.0 &   ISO$-$SWS1  &   81002721    \\
\hline
GZ Peg  &   735 &   1100    &   965 &   88.6    &   78.9    &   62.4    &   56.8    &   40.9    &   21.0    &   ISO$-$SWS1  &   37600306    \\
ST Her  &   804 &   1162    &   1098    &   166 &   186 &   200 &   187 &   149 &   104 &   ISO$-$SWS1  &   41901305    \\
T Cet   &   1009    &   1596    &   1403    &   172 &   163 &   172 &   171 &   134 &   83.0    &   ISO$-$SWS1  &   55502308 $-$ 37801819   \\
NO Aur  &   226 &   362 &   273 &   33.2    &   41.6    &   46.3    &   36.1    &   23.3    &   17.2    &   ISO$-$SWS1  &   86603434    \\
RX Lac  &   465 &   737 &   681 &   91.0    &   85.4    &   81.9    &   68.6    &   50.3    &   28.8    &   ISO$-$SWS1  &   78200427    \\
\hline
R And   &   247 &   506 &   596 &   193 &   264 &   248 &   210 &   176 &   135 &   ISO$-$SWS1  &   40201723    \\
HR Peg  &   191 &   327 &   256 &   27.0    &   24.4    &   20.9    &   20.0    &   12.6    &   7.8 &   ISO$-$SWS1  &   37401910    \\
OP Her  &   416 &   731 &   556 &   63.5    &   58.1    &   46.4    &   37.8    &   30.9    &   16.0    &   ISO$-$SWS1  &   77800625    \\
\hline
LY Mus  &   221 &   334 &   300 &   28.4    &   24.7    &   19.7    &   16.4    &   12.9    &   7.6 &   ISO$-$SWS1  &   13201304    \\
CSS 1023    &   33.2    &   51.7    &   39.9    &   2.3 &   2.2 &   1.5 &   1.4 &   0.83    &   0.46    &   ISO$-$SWS1  &   14100603    \\
\hline
S Lyr   &   8.2 &   13.2    &   17.9    &   20.0    &   23.2    &   25.5    &   24.4    &   22.1    &   15.7    &   ISO$-$SWS1  &   52000546    \\
\hline
II Lup  &   3.7 &   25.5    &   99.0    &   860 &   852 &   860 &   681 &   506 &   391 &   ISO$-$SWS6  &   29700401    \\
\hline      \hline
\end{tabular}
\end{table*}

\begin{table*}[t!]
\caption{Sample A $-$ Third part. The indication I. $-$ E.
(Intrinsic $-$ Extrinsic) is given according to the suggestions
from \citet{vaneck00} and \citet{yang}. In discordant cases we
prefer the choice by \citet{vaneck00} and show the one of
\citet{yang} in parenthesis. For few sources, for which neither
study offers a suggestion, we infer that they are extrinsic from
their low intrinsic Luminosity: in the tables they are underlined.}             
\label{table:3}      
\centering                          
\begin{tabular}{c c c c c c c c}        
\hline              \hline
Source  &   Var. Type   &   Period  &   Distance    &   Min. $-$ Max.   &   Ref.    &   Bol. Magnitudes &   I. $-$ E.   \\
name    &   (GCVS)  &   (GCVS)  &   (kpc)   &   (kpc)   &   Distance    &   ISO Integration &       \\
\hline              \hline
S Cas   &   M   &   612.43  &   0.85    &   $-$ &   P$-$L / this paper  &   $-$5.71 &   I   \\
W Aql   &   M   &   490.43  &   0.34    &   $-$ &   P$-$L / this paper  &   $-$5.44 &   I   \\
R Cyg   &   M   &   426.45  &   0.55    &   $-$ &   P$-$L / this paper  &   $-$5.42 &   $-$ \\
chi Cyg &   M   &   408.05  &   0.18    &   0.15 $-$ 0.22   &   Hip. / \cite{vleu2007}  &   $-$5.39 &   I   \\
W Cet   &   M   &   351.31  &   0.83    &   $-$ &   P$-$L / this paper  &   $-$5.18 &   I   \\
pi1 Gru &   SRB &   150 &   0.16    &   0.15 $-$ 0.19   &   Hip. / \cite{vleu2007}  &   $-$5.75 &   I   \\
AA Cyg  &   SRB &   212.7   &   $-$ &   $-$ &   $-$ &   $-$ &   I   \\
RZ Sgr  &   SRB &   223.2   &   $-$ &   $-$ &   $-$ &   $-$ &   I   \\
WX Cam  &   LB  &   $-$ &   $-$ &   $-$ &   $-$ &   $-$ &   I   \\
\hline
GZ Peg  &   SRA &   92.66   &   0.24    &   0.22 $-$ 0.26   &   Hip. / \cite{vleu2007}  &   $-$5.02 &   E   \\
ST Her  &   SRB &   148 &   0.30    &   0.25 $-$ 0.36   &   Hip. / \cite{vleu2007}  &   $-$5.64 &   I   \\
T Cet   &   SRC &   158.9   &   0.27    &   0.24 $-$ 0.31   &   Hip. / \cite{vleu2007}  &   $-$5.63 &   I   \\
NO Aur  &   LC  &   $-$ &   0.60    &   0.47 $-$ 0.83   &   Hip. / \cite{vleu2007}  &   $-$5.73 &   I   \\
RX Lac  &   SRB &   650 &   $-$ &   $-$ &   $-$ &   $-$ &   I   \\
\hline
R And   &   M   &   409.33  &   0.41    &   $-$ &   P$-$L / this paper  &   $-$5.19 &   I   \\
HR Peg  &   SRB &   50  &   0.41    &   0.36 $-$ 0.50   &   Hip. / \cite{vleu2007}  &   $-$4.75 &   I   \\
OP Her  &   SRB &   120.5   &   0.30    &   0.27 $-$ 0.32   &   Hip. / \cite{vleu2007}  &   $-$4.92 &   $-$ \\
\hline
LY Mus  &   LB  &   $-$ &   0.29    &   0.26 $-$ 0.33   &   Hip. / \cite{vleu2007}  &   $-$4.12 &   E   \\
CSS 1023    &   $-$ &   $-$ &   $-$ &   $-$ &   $-$ &   $-$ &   E   \\
\hline
S Lyr   &   M   &   438.4   &   2.27    &   $-$ &   P$-$L / this paper  &   $-$5.50 &   I   \\
\hline
II Lup  &   M   &   $-$ &   0.59    &   $-$ &   \citet{groenewegen02b}  &   $-$4.82 &   $-$ \\
\hline              \hline
\end{tabular}
\end{table*}

\begin{table*}
\caption{Sample B $-$ First part. Suggestions from \citet{sl86}
for Spectral Type: \emph{1} RS Cnc: M6eIIIaS. \emph{2} V1743 Cyg: M5IIIaS. \emph{3} V1981 Cyg: M4IIIaS.}             
\label{table:4}      
\centering                          
\begin{tabular}{c c c c c c}        
\hline \hline
IRAS    &   Other   &   Stephenson  &   Coordinates &   Spectral Type   &   Var. Type   \\
name    &   name    &   name    &   ICRS    &       &   (GCVS)  \\
\hline \hline
05199$-$0842    &   \object{V1261 Ori}  &   CSS 133 &   05 22 18.6453 $-$08 39 58.034   &   S…  &   Algol Type  \\
\hline
04497+1410  &   \object{omi Ori}    &   CSS 114 &   04 52 31.9621 +14 15 02.311 &   M3.2IIIaS   &   SRB \\
10226+0902  &   \object{DE Leo} &   $-$ &   10 25 15.1951 +08 47 05.441 &   M2IIIabS    &   SRB \\
07245+4605  &   \object{Y Lyn}  &   CSS 347 &   07 28 11.6109 +45 59 26.207 &   M6SIb$-$II  &   SRC \\
07392+1419  &   \object{NZ Gem} &   CSS 382 &   07 42 03.2185 +14 12 30.612 &   M3II$-$IIIS &   SR  \\
\hline
06457+0535  &   \object{V613 Mon}   &   CSS 260 &   06 48 22.2963 +05 32 30.050 &   M2/S5,1 &   SRB \\
09076+3110  &   \object{RS Cnc} &   CSS 589 &   09 10 38.7990 +30 57 47.300 &   M6eIb$-$II/S \footnote{}    &   SRC \\
03377+6303  &   \object{BD Cam} &   CSS 79  &   03 42 09.3250 +63 13 00.501 &   S5,3/M4III  &   LB  \\
07095+6853  &   \object{AA Cam} &   CSS 312 &   07 14 52.0703 +68 48 15.380 &   M5/S    &   LB  \\
13079$-$8931    &   \object{BQ Oct} &   CSS 804 &   14 35 29.5001 $-$89 46 18.182   &   M4III/S5,1  &   LB  \\
\hline
12272$-$4127    &   \object{V928 Cen}   &   CSS 796 &   12 29 57.8871 $-$41 44 09.242   &   M2II$-$III  &   SRB \\
19323+4909  &   \object{V1743 Cyg}  &   $-$ &   19 33 41.6068 +49 15 44.347 &   M4.5III \footnote{} &   SRB \\
$-$ &   \object{V1981 Cyg}  &   $-$ &   21 02 24.1993 +44 47 27.528     &   M4s... \footnote{}  &   SRB \\
08214$-$3807    &   \object{V436 Pup}   &   CSS 500 &   08 23 16.9344 $-$38 17 09.884   &   M1III   &   LB  \\
$-$ &   \object{V2141 Cyg}  &   CSS 1254    &   20 57 53.1771 +44 47 17.336 &   M1  &   LB  \\
12106$-$3350    &   \object{V335 Hya}   &   $-$ &   12 13 12.9423 $-$34 07 30.981   &   M4III   &   LB  \\
14510$-$6052    &   \object{CR Cir} &   CSS 867 &   14 54 56.9389 $-$61 04 33.027   &   M2/M3II &   LC  \\
16418$-$1359    &   $-$ &   \object{CSS 937}    &   16 44 42.1936 $-$14 04 48.553   &   M1III   &   $-$ \\
\hline
13136$-$4426    &   \object{UY Cen} &   CSS 816 &   13 16 31.8300 $-$44 42 15.741   &   SC  &   SR  \\
\hline
16425$-$1902    &   $-$ &   \object{CSS 938}    &   16 45 30.1769 $-$19 08 12.939   &   K5II    &   $-$ \\
20076+3331  &   $-$ &   \object{CSS 1194}   &   20 09 32.9873 +33 40 53.851 &   K5III   &   $-$ \\
\hline \hline
\end{tabular}
\end{table*}

\begin{table*}
\caption{Sample B $-$ Second part.}             
\label{table:5}      
\centering                          
\begin{tabular}{c c c c c c c c c c c}        
\hline      \hline
Source  &   J   &   H   &   K   &   [8.8]   &   [9.8]   &   [11.7]  &   [12.5]  &   D   &   E   &   Mid$-$IR Data Origin    \\
name    &   [Jy]    &   [Jy]    &   [Jy]    &   [Jy]    &   [Jy]    &   [Jy]    &   [Jy]    &   [Jy]    &   [Jy]    &       \\
\hline      \hline
V1261 Ori   &   73.8    &   111 &   93.1    &   16.4    &   16.5    &   19.6    &   21.3    &   $-$ &   $-$ &   IRAS$-$LRS  \\
\hline
omi Ori &   1022    &   1587    &   1227    &   103 &   82.4    &   64.8    &   60.9    &   $-$ &   $-$ &   IRAS$-$LRS  \\
DE Leo  &   178 &   234 &   184 &   34.0    &   30.7    &   34.4    &   36.4    &   $-$ &   $-$ &   IRAS$-$LRS  \\
Y Lyn   &   876 &   1448    &   1256    &   132 &   150 &   121 &   107 &   $-$ &   $-$ &   IRAS$-$LRS  \\
NZ Gem  &   369 &   566 &   399 &   38.9    &   34.5    &   32.6    &   32.5    &   $-$ &   $-$ &   IRAS$-$LRS  \\
\hline
V613 Mon    &   49.6    &   74.3    &   68.9    &   5.2 &   $-$ &   $-$ &   3.3 &   2.0 &   $-$ &   MSX \\
RS Cnc  &   3065    &   4324    &   3742    &   512 &   693 &   493 &   436 &   $-$ &   $-$ &   TIRCAM2 \\
BD Cam  &   439 &   630 &   521 &   62.3    &   52.5    &   43.6    &   41.2    &   $-$ &   $-$ &   IRAS$-$LRS  \\
AA Cam  &   146 &   220 &   185 &   $-$ &   $-$ &   $-$ &   $-$ &   $-$ &   $-$ &   $-$ \\
BQ Oct  &   139 &   196 &   170 &   24.8    &   22.5    &   23.5    &   23.6    &   $-$ &   $-$ &   IRAS$-$LRS  \\
\hline
V928 Cen    &   196 &   267 &   225 &   25.4    &   21.9    &   20.3    &   20.5    &   $-$ &   $-$ &   IRAS$-$LRS  \\
V1743 Cyg   &   259 &   421 &   335 &   40.4    &   36.4    &   34.5    &   34.2    &   $-$ &   $-$ &   IRAS$-$LRS  \\
V1981 Cyg   &   197 &   279 &   245 &   28.9    &   $-$ &   $-$ &   16.3    &   10.8    &   4.7 &   MSX \\
V436 Pup    &   139 &   198 &   165 &   17.1    &   $-$ &   $-$ &   10.3    &   7.0 &   2.7 &   MSX \\
V2141 Cyg   &   147 &   216 &   195 &   20.8    &   $-$ &   $-$ &   12.2    &   8.2 &   3.5 &   MSX \\
V335 Hya    &   611 &   946 &   833 &   88.6    &   78.7    &   67.4    &   63.3    &   $-$ &   $-$ &   IRAS$-$LRS  \\
CR Cir  &   75.1    &   117 &   99.6    &   9.9 &   $-$ &   $-$ &   5.8 &   3.8 &   $-$ &   MSX \\
CSS 937 &   104 &   169 &   151 &   15.7    &   15.9    &   20.1    &   21.6    &   $-$ &   $-$ &   IRAS$-$LRS  \\
\hline
UY Cen  &   184 &   360 &   350 &   66.7    &   62.9    &   61.4    &   55.6    &   $-$ &   $-$ &   IRAS$-$LRS  \\
\hline
CSS 938 &   142 &   200 &   175 &   20.4    &   19.7    &   21.4    &   23.2    &   $-$ &   $-$ &   IRAS$-$LRS  \\
CSS 1194    &   31.1    &   41.5    &   37.0    &   2.9 &   $-$ &   $-$ &   2.4 &   0.96    &   $-$ &   MSX \\
\hline      \hline
\end{tabular}
\end{table*}

\begin{table*}
\caption{Sample B $-$ Third part.}             
\label{table:6}      
\centering                          
\begin{tabular}{c c c c c c c c}        
\hline              \hline
Source  &   Var. Type   &   Period  &   Distance    &   Min. $-$ Max.   &   Ref.    &   Bol. Magnitudes &   I. $-$ E.   \\
name    &   (GCVS)  &   (GCVS)  &   (kpc)   &   (kpc)   &   Distance    &   Bol. Corrections    &       \\
\hline              \hline
V1261 Ori   &   Algol Type  &   $-$ &   0.29    &   0.23 $-$ 0.38   &   Hip. / \cite{vleu2007}  &   $-$2.71 &   E   \\
\hline
omi Ori &   SRB &   30  &   0.20    &   0.17 $-$ 0.23   &   Hip. / \cite{vleu2007}  &   $-$4.89 &   I   \\
DE Leo  &   SRB &   $-$ &   0.31    &   0.27 $-$ 0.37   &   Hip. / \cite{vleu2007}  &   $-$3.60 &   \textbf{$\underline{E}$}    \\
Y Lyn   &   SRC &   110 &   0.25    &   0.20 $-$ 0.33   &   Hip. / \cite{vleu2007}  &   $-$5.33 &   I   \\
NZ Gem  &   SR  &   $-$ &   0.39    &   0.32 $-$ 0.51   &   Hip. / \cite{vleu2007}  &   $-$5.06 &   E   \\
\hline
V613 Mon    &   SRB &   $-$ &   0.50    &   0.35 $-$ 0.84   &   Hip. / \cite{vleu2007}  &   $-$3.76 &   E   \\
RS Cnc  &   SRC &   120 &   0.14    &   0.13 $-$ 0.15   &   Hip. / \cite{vleu2007}  &   $-$5.21 &   I   \\
BD Cam  &   LB  &   $-$ &   0.16    &   0.15 $-$ 0.17   &   Hip. / \cite{vleu2007}  &   $-$3.40 &   E   \\
AA Cam  &   LB  &   $-$ &   0.78    &   0.50 $-$ 1.82   &   Hip. / \cite{vleu2007}  &   $-$ &   I   \\
BQ Oct  &   LB  &   $-$ &   0.49    &   0.41 $-$ 0.60   &   Hip. / \cite{vleu2007}  &   $-$4.55 &   I   \\
\hline
V928 Cen    &   SRB &   $-$ &   0.23    &   0.21 $-$ 0.25   &   Hip. / \cite{vleu2007}  &   $-$3.27 &   E   \\
V1743 Cyg   &   SRB &   40  &   0.41    &   0.38 $-$ 0.44   &   Hip. / \cite{vleu2007}  &   $-$4.94 &   $-$ \\
V1981 Cyg   &   SRB &   $-$ &   0.30    &   0.27 $-$ 0.33   &   Hip. / \cite{vleu2007}  &   $-$3.96 &   \textbf{$\underline{E}$}    \\
V436 Pup    &   LB  &   $-$ &   0.33    &   0.30 $-$ 0.38   &   Hip. / \cite{vleu2007}  &   $-$3.75 &   E   \\
V2141 Cyg   &   LB  &   $-$ &   0.38    &   0.31 $-$ 0.51   &   Hip. / \cite{vleu2007}  &   $-$4.24 &   \textbf{$\underline{E}$}    \\
V335 Hya    &   LB  &   $-$ &   0.36    &   0.31 $-$ 0.43   &   Hip. / \cite{vleu2007}  &   $-$5.68 &   $-$ \\
CR Cir  &   LC  &   $-$ &   0.31    &   0.25 $-$ 0.42   &   Hip. / \cite{vleu2007}  &   $-$3.08 &   E   \\
CSS 937 &   $-$ &   $-$ &   0.42    &   0.31 $-$ 0.66   &   Hip. / \cite{vleu2007}  &   $-$4.11 &   E   \\
\hline
UY Cen  &   SR  &   114.6   &   0.69    &   0.47 $-$ 1.33   &   Hip. / \cite{vleu2007}  &   $-$6.05 &   I   \\
\hline
CSS 938 &   $-$ &   $-$ &   0.25    &   0.20 $-$ 0.33   &   Hip. / \cite{vleu2007}  &   $-$3.15 &   E   \\
CSS 1194    &   $-$ &   $-$ &   0.36    &   0.30 $-$ 0.45   &   Hip. / \cite{vleu2007}  &   $-$2.34 &   \textbf{$\underline{E}$}    \\
\hline              \hline
\end{tabular}
\end{table*}

Due to the above phenomena, the atmospheres of AGB stars are
characterized by an increasing enrichment of $^{12}$C (up to C/O $>$
1 by number, in which case we speak of C stars) and of s-process
nuclei, in particular revealing the recent nucleosynthesis through
the short-lived $^{99}$Tc \citep{merrill}. Sometimes Tc itself
offers actually the only real evidence of ongoing s-processing in
stars that do not show other remarkable chemical anomalies
\citep{utt07}. When instead the third dredge-up process is efficient
enough, changes in the photospheric abundances of other s-elements
begin to occur, first of all for Zr, which has various isotopes on
the main s-process path. In such cases the appearance of ZrO bands
in the spectra (at wavelengths 464.1, 462.0, 530.4, 537.9, and 555.1
nm) tells us that the star, although still richer in O than in C, is
mixing to the surface the products of shell-He burning. The cool
giants presenting these signatures are called MS and S stars (the
second group showing more prominent features). They also have a C/O
abundance ratio by number higher than in the Sun, but lower than
unity. There is still some confusion about the exact values of the
C/O ratios in MS and S stars, which is mainly induced by the
remaining uncertainty in the calibrating solar oxygen abundance. If
one excludes for this calibration the recent, still debated
suggestion \citep{apla} and adopts instead the previous more
traditional reference \citep{ag89}, then MS stars are found
typically at C/O = 0.5 to 0.7, and S stars are found above this
range and up to more that 0.95. Around the border between O-rich and
C-rich giants, the so-called SC stars represent a rare, but
important, transition group.

In some cases, red giants showing s-element enhancement are Tc-poor.
This is a clear indication that a sufficiently long time interval
has passed since the production of neutron capture nuclei, so that
Tc has decayed. In general, this is the case when the
nucleosynthesis phenomena occurred not in the same star we see
today, but in a more massive companion, which is now evolved to the
white dwarf stage, and whose mass loss enriched in the past the
photosphere of the observed object \citep{busso01}. In these cases
we speak more properly of an {\it extrinsic} AGB star
\citep{smilamb}.

AGB stars lose mass very effectively, and their winds replenish
the Interstellar Medium (ISM) guaranteeing up to 70\% of the mass
return from stars \citep{sedlmayr94}. Before being dispersed over
the Galaxy, the material thus lost forms cool envelopes
\citep{winters03} where dust grains condense \citep{carciofi04}.
These solid particles carry the elemental and isotopic composition
generated by AGB nucleosynthesis; they have been found in ancient
meteorites offering the possibility of high precision isotopic
abundance measurements on matter coming from circumstellar
environments \citep[e.g.][]{zinner00}.

The cool AGB photospheres radiate most of their flux at
red-infrared wavelengths. The infrared component of the Spectral
Energy Distribution (SED) grows in importance (and in average
wavelength) as far as the evolution proceeds, because of the
increased extinction of the photospheric flux operated by dust,
which then re-radiates at long wavelengths \citep[see
e.g.][]{habing96}. This correlation between extinction and
evolutionary stage is however confused by the stars switching from
Semiregular to Mira-type surface variability, which fact modulates
the mass loss efficiency and hence the extinction properties. Due
to these complicacies, large surveys of infrared (IR) observations
play a fundamental role in studying luminosities and mass loss
rates of AGB stars and in disentangling the variability and
evolution effects \citep[see
e.g.][]{woodcohen,lebertre01,lebertre03,groenewegen02b,cioni03,omont03,olofsson03}.

Longstanding efforts have been devoted to describe the mass loss
mechanisms, either with phenomenological models or with
sophisticated hydrodynamical approaches
\citep{salpeter,knappmorris,winters03,wachter02,sandin3a,sandin3b}.
Despite this, our quantitative knowledge of AGB winds is still
poor and forces us to adopt parametric treatments, where
observations play a crucial role in fixing the (otherwise free)
parameters \citep{wood03,olivier,wood04,andersen03}.

Similar problems affect the estimates of the stellar luminosity.
Observations are hampered by the difficulties of measuring the
distances for single, often obscured objects like AGB stars. On
the other hand, the luminosities derived from full stellar
evolutionary models are affected by the uncertainties in the
choice of the mixing parameters (in particular of the extension of
convective overshoot) and of surface atomic and molecular
opacities \citep{marigo}. Models adopting large overshoot
parameters \citep[see e.g.][]{izz07} derive large values for the
mass dredged-up after each thermal pulse, thus obtaining the
surface enrichment in C and s-elements earlier and at a lower
luminosity than models based on the Schwarzschild's criterion can
do.

\begin{table*}[t!]
\caption{Sample C $-$ First part.}             
\label{table:7}      
\centering                          
\begin{tabular}{c c c c c c}        
\hline \hline
IRAS    &   Other   &   Stephenson  &   Coordinates &   Spectral Type   &   Var. Type   \\
name    &   name    &   name    &   ICRS    &       &   (GCVS)  \\
\hline \hline
04352+6602  &   \object{T Cam}  &   CSS 103 &   04 40 08.8768 +66 08 48.654     &   S4,7e$-$S8.5,8e &   M   \\
06571+5524  &   \object{R Lyn}  &   CSS 283 &   07 01 18.0093 +55 19 49.766     &   S2.5,5e$-$S6,8e:    &   M   \\
07043+2246  &   \object{R Gem}  &   CSS 307 &   07 07 21.2744 +22 42 12.736 &   S2,9e$-$S8,9e(Tc)   &   M   \\
07092+0735  &   \object{WX CMi} &   CSS 316 &   07 11 57.45 +07 29 59.3 &   Se  &   M   \\
12417+6121  &   \object{S UMa}  &   CSS 803 &   12 43 56.676 +61 05 35.51   &   S0,9e$-$S5,9e   &   M   \\
15030$-$4116    &   \object{GI Lup} &   CSS 872 &   15 06 16.31 $-$41 28 14.1   &   S7,8e   &   M   \\
23554+5612  &   \object{WY Cas} &   CSS 1345    &   23 58 01.30 +56 29 13.5 &   S6,5pe  &   M   \\
00135+4644  &   \object{X And}  &   CSS 6   &   00 16 09.57 +47 00 44.8 &   S2,9e$-$S5,5e   &   M   \\
00435+4758  &   \object{U Cas}  &   CSS 12  &   00 46 21.371 +48 14 38.72   &   S3,5e$-$S8,6e   &   M   \\
06062+2830  &   \object{GH Aur} &   CSS 191 &   06 09 27.71 +28 29 43.4 &   S   &   M   \\
07197$-$1451    &   \object{TT CMa} &   CSS 341 &   07 22 02.00 $-$14 56 56.7   &   S   &   M   \\
07584$-$2051    &   \object{EX Pup} &   CSS 443 &   08 00 38.31 $-$20 59 35.3   &   S2,4e   &   M   \\
11179$-$6135    &   \object{RY Car} &   CSS 742 &   11 20 11.39 $-$61 52 16.8   &   S7,8e   &   M   \\
13226$-$6302    &   \object{NZ Cen} &   CSS 820 &   13 26 02.52 $-$63 18 28.5   &   Se  &   M   \\
14212+8403  &   \object{R Cam}  &   CSS 856 &   14 17 51.0439 +83 49 53.861     &   S2,8e$-$S8,7e   &   M   \\
17478$-$2957    &   \object{V762 Sgr}   &   CSS 1001    &   17 51 04.04 $-$29 58 30.9   &   S6,4    &   M   \\
17490$-$3502    &   \object{V407 Sco}   &   CSS 1004    &   17 52 25.53 $-$35 03 17.5   &   Se  &   M   \\
19166+0318  &   \object{ER Aql} &   CSS 1121    &   19 19 06.99 +03 24 05.2 &   S   &   M   \\
23376+6304  &   \object{V441 Cas}   &   CSS 1338    &   23 39 58.92 +63 20 55.1 &   S   &   M   \\
23489+6235  &   \object{EO Cas} &   CSS 1342    &   23 51 27.30 +62 51 47.0 &   Se  &   M   \\
00001+4826  &   \object{IW Cas} &   CSS 1347    &   00 02 44.22 +48 42 50.9 &   S4.5,9e &   M   \\
$-$ &   $-$ &   \object{CSS2 10}    &   02 51 33.00 +57 50 34.5 &   S   &   M   \\
\hline
10237$-$6135    &   \object{AU Car} &   CSS 679 &   10 25 29.74 $-$61 50 59.1   &   MS  &   M   \\
10349$-$6203    &   \object{RX Car} &   CSS 690 &   10 36 45.82 $-$62 19 16.8   &   MS  &   M   \\
\hline
02143+4404  &   \object{W And}  &   CSS 49  &   02 17 32.9606 +44 18 17.766 &   S6,1e$-$S9,2e/M4$-$M1   &   M   \\
07149+0111  &   \object{RR Mon} &   CSS 326 &   07 17 31.54 +01 05 41.5 &   S7,2e$-$S8,2e/M6$-$10   &   M   \\
07545$-$4400    &   \object{SU Pup} &   CSS 436 &   07 56 12.0813 $-$44 08 33.254   &   M/S4,2e &   M   \\
09338$-$5349    &   \object{UU Vel} &   CSS 614 &   09 35 33.21 $-$54 03 25.9   &   M2e/S7,8e   &   M   \\
17001$-$3651    &   \object{RT Sco} &   CSS 954 &   17 03 32.56 $-$36 55 13.7   &   S7,2/M6e$-$M7e  &   M   \\
20213+0047  &   \object{V865 Aql}   &   CSS 1211    &   20 23 54.6422 +00 56 44.794 &   M6$-$M7/S7,5e:  &   M   \\
00445+3224  &   \object{RW And} &   CSS 14  &   00 47 18.92 +32 41 08.6 &   M5e$-$M10e/S6,2e    &   M   \\
07103$-$0258    &   \object{AK Mon} &   CSS 319 &   07 12 49.91 $-$03 03 29.0   &   M5/S5,1 &   M   \\
17521$-$2907    &   \object{V745 Sgr}   &   CSS 1007    &   17 55 19.00 $-$29 07 54.4   &   Se/M    &   M   \\
20044+5750  &   \object{S Cyg}  &   CSS 1191    &   20 05 29.85 +57 59 09.1 &   S2.5,1e/M3.5$-$M7e  &   M   \\
20369+3742  &   \object{FF Cyg} &   CSS 1232    &   20 38 51.71 +37 53 23.2 &   S6,8e/M4e   &   M   \\
\hline
03499+4730  &   \object{FG Per} &   CSS 85  &   03 53 30.2 +47 39 04    &   M9  &   M   \\
\hline
13163$-$6031    &   \object{TT Cen} &   CSS 817 &   13 19 35.016 $-$60 46 46.26 &   CSe &   M   \\
18586$-$1249    &   \object{ST Sgr} &   CSS 1096    &   19 01 29.20 $-$12 45 34.0   &   C4,3e$-$S9,5e   &   M   \\
21540+4806  &   \object{LX Cyg} &   CSS 1286    &   21 55 57.03 +48 20 52.6 &   SC3e$-$S5,5e:   &   M   \\
\hline
18575$-$0139    &   \object{VX Aql} &   CSS 1093    &   19 00 09.61 $-$01 34 56.8   &   C9,1p/M0ep  &   M   \\
\hline
01097+6154  &   \object{V418 Cas}   &   CSS 23  &   01 12 59.89 +62 10 47.6 &   $-$ &   M   \\
\hline \hline
\end{tabular}
\end{table*}

\begin{table*}[t!]
\caption{Sample C $-$ Second part.}             
\label{table:8}      
\centering                          
\begin{tabular}{c c c c c c c c c c c}        
\hline      \hline
Source  &   J   &   H   &   K   &   [8.8]   &   [9.8]   &   [11.7]  &   [12.5]  &   D   &   E   &   Mid$-$IR Data Origin    \\
name    &   [Jy]    &   [Jy]    &   [Jy]    &   [Jy]    &   [Jy]    &   [Jy]    &   [Jy]    &   [Jy]    &   [Jy]    &       \\
\hline      \hline
T Cam   &   152 &   306 &   315 &   62.1    &   53.4    &   48.8    &   47.1    &   $-$ &   $-$ &   IRAS$-$LRS  \\
R Lyn   &   51.7    &   77.7    &   91.1    &   29.9    &   29.9    &   29.2    &   28.5    &   $-$ &   $-$ &   IRAS$-$LRS  \\
R Gem   &   155 &   226 &   174 &   24.1    &   $-$ &   $-$ &   21.4    &   16.3    &   7.5 &   MSX \\
WX CMi  &   8.9 &   13.8    &   17.3    &   8.3 &   $-$ &   $-$ &   9.2 &   6.3 &   4.1 &   MSX \\
S UMa   &   26.3    &   43.4    &   41.4    &   4.5 &   $-$ &   2.8 &   3.1 &   $-$ &   $-$ &   TIRCAM2 \\
GI Lup  &   61.7    &   115 &   128 &   $-$ &   $-$ &   $-$ &   $-$ &   $-$ &   $-$ &   $-$ \\
WY Cas  &   64.5    &   94.4    &   120 &   48.8    &   55.6    &   58.2    &   55.0    &   $-$ &   $-$ &   IRAS$-$LRS  \\
X And   &   18.1    &   37.3    &   43.3    &   23.6    &   23.6    &   24.9    &   26.6    &   $-$ &   $-$ &   IRAS$-$LRS  \\
U Cas   &   29.6    &   45.0    &   47.1    &   15.2    &   15.6    &   19.4    &   20.3    &   $-$ &   $-$ &   IRAS$-$LRS  \\
GH Aur  &   4.4 &   7.8 &   10.2    &   1.5 &   $-$ &   $-$ &   1.8 &   $-$ &   $-$ &   MSX \\
TT CMa  &   29.6    &   50.1    &   54.3    &   12.3    &   $-$ &   $-$ &   14.1    &   10.8    &   7.7 &   MSX \\
EX Pup  &   3.9 &   5.9 &   5.8 &   0.54    &   $-$ &   $-$ &   $-$ &   $-$ &   $-$ &   MSX \\
RY Car  &   8.3 &   18.2    &   25.1    &   11.0    &   $-$ &   $-$ &   8.9 &   5.6 &   3.2 &   MSX \\
NZ Cen  &   8.3 &   15.7    &   17.2    &   9.1 &   $-$ &   $-$ &   9.3 &   6.3 &   5.2 &   MSX \\
R Cam   &   47.2    &   70.8    &   69.5    &   7.1 &   $-$ &   $-$ &   $-$ &   2.8 &   $-$ &   MSX \\
V762 Sgr    &   15.0    &   35.9    &   44.5    &   13.5    &   $-$ &   $-$ &   14.4    &   9.3 &   5.1 &   MSX \\
V407 Sco    &   9.8 &   18.1    &   19.4    &   6.1 &   $-$ &   $-$ &   5.6 &   3.8 &   $-$ &   MSX \\
ER Aql  &   34.4    &   61.5    &   62.3    &   10.1    &   $-$ &   $-$ &   6.9 &   4.7 &   $-$ &   MSX \\
V441 Cas    &   5.9 &   11.6    &   14.3    &   2.0 &   $-$ &   $-$ &   1.5 &   $-$ &   $-$ &   MSX \\
EO Cas  &   11.8    &   21.4    &   30.9    &   9.3 &   $-$ &   $-$ &   7.5 &   4.8 &   3.0 &   MSX \\
IW Cas  &   23.3    &   47.4    &   50.9    &   52.9    &   58.5    &   61.8    &   60.5    &   $-$ &   $-$ &   IRAS$-$LRS  \\
CSS2 10 &   1.3 &   2.8 &   3.1 &   0.36    &   $-$ &   $-$ &   $-$ &   $-$ &   $-$ &   MSX \\
\hline
AU Car  &   5.9 &   9.0 &   10.3    &   2.1 &   $-$ &   $-$ &   $-$ &   $-$ &   $-$ &   MSX \\
RX Car  &   5.0 &   7.7 &   8.9 &   2.1 &   $-$ &   $-$ &   1.7 &   $-$ &   $-$ &   MSX \\
\hline
W And   &   368 &   643 &   591 &   185 &   198 &   163 &   143 &   $-$ &   $-$ &   IRAS$-$LRS  \\
RR Mon  &   24.1    &   42.5    &   52.8    &   21.3    &   $-$ &   $-$ &   18.0    &   12.7    &   7.2 &   MSX \\
SU Pup  &   28.5    &   43.8    &   45.5    &   22.3    &   26.6    &   29.1    &   29.3    &   $-$ &   $-$ &   IRAS$-$LRS  \\
UU Vel  &   23.5    &   46.9    &   51.6    &   13.6    &   $-$ &   $-$ &   11.5    &   7.5 &   4.1 &   MSX \\
RT Sco  &   285 &   460 &   512 &   162 &   $-$ &   $-$ &   163 &   108 &   67.4    &   MSX \\
V865 Aql    &   126 &   191 &   199 &   36.5    &   33.8    &   35.2    &   35.8    &   $-$ &   $-$ &   IRAS$-$LRS  \\
RW And  &   95.9    &   132 &   131 &   48.8    &   52.8    &   53.0    &   50.1    &   $-$ &   $-$ &   IRAS$-$LRS  \\
AK Mon  &   9.3 &   13.1    &   14.7    &   3.1 &   $-$ &   $-$ &   2.8 &   2.2 &   $-$ &   MSX \\
V745 Sgr    &   38.2    &   75.8    &   82.5    &   18.5    &   $-$ &   $-$ &   16.1    &   11.3    &   7.4 &   MSX \\
S Cyg   &   11.8    &   14.0    &   16.5    &   2.0 &   $-$ &   $-$ &   1.5 &   0.74    &   $-$ &   MSX \\
FF Cyg  &   37.8    &   65.8    &   82.0    &   8.5 &   $-$ &   $-$ &   6.4 &   4.9 &   $-$ &   MSX \\
\hline
FG Per  &   3.2 &   6.0 &   6.8 &   1.1 &   $-$ &   $-$ &   $-$ &   $-$ &   $-$ &   MSX \\
\hline
TT Cen  &   19.5    &   48.9    &   54.7    &   15.5    &   $-$ &   $-$ &   16.3    &   10.1    &   8.0 &   MSX \\
ST Sgr  &   97.7    &   148 &   149 &   61.6    &   63.2    &   61.1    &   56.4    &   $-$ &   $-$ &   IRAS$-$LRS  \\
LX Cyg  &   14.2    &   28.0    &   44.9    &   10.2    &   $-$ &   $-$ &   9.4 &   5.1 &   $-$ &   MSX \\
\hline
VX Aql  &   13.5    &   33.6    &   37.7    &   9.3 &   $-$ &   $-$ &   10.6    &   6.8 &   3.5 &   MSX \\
\hline
V418 Cas    &   11.0    &   16.0    &   22.8    &   12.5    &   $-$ &   $-$ &   12.2    &   7.9 &   5.5 &   MSX \\
\hline      \hline
\end{tabular}
\end{table*}

\begin{table*}[t!]
\caption{Sample C $-$ Third part.}             
\label{table:9}      
\centering                          
\begin{tabular}{c c c c c c c}        
\hline              \hline
Source  &   Var. Type   &   Period  &   Distance    &   Ref.    &   Bol. Magnitudes &   I. $-$ E.   \\
name    &   (GCVS)  &   (GCVS)  &   (kpc)   &   Distance    &   P$-$L Method    &       \\
\hline              \hline
T Cam   &   M   &   373.2   &   0.50    &   P$-$L / this paper  &   $-$5.22 &   I   \\
R Lyn   &   M   &   378.75  &   0.95    &   P$-$L / this paper  &   $-$5.19 &   I   \\
R Gem   &   M   &   369.91  &   0.66    &   P$-$L / this paper  &   $-$5.23 &   I   \\
WX CMi  &   M   &   420.1   &   2.35    &   P$-$L / this paper  &   $-$5.31 &   $-$ \\
S UMa   &   M   &   225.87  &   0.96    &   P$-$L / this paper  &   $-$4.56 &   I   \\
GI Lup  &   M   &   326.2   &   0.80    &   P$-$L / this paper  &   $-$ &   I   \\
WY Cas  &   M   &   476.56  &   0.97    &   P$-$L / this paper  &   $-$5.50 &   I   \\
X And   &   M   &   346.18  &   1.31    &   P$-$L / this paper  &   $-$5.03 &   I   \\
U Cas   &   M   &   277.19  &   1.08    &   P$-$L / this paper  &   $-$4.74 &   I   \\
GH Aur  &   M   &   349 &   2.64    &   P$-$L / this paper  &   $-$5.13 &   I   \\
TT CMa  &   M   &   314 &   1.08    &   P$-$L / this paper  &   $-$4.95 &   I   \\
EX Pup  &   M   &   289 &   3.02    &   P$-$L / this paper  &   $-$4.92 &   $-$ \\
RY Car  &   M   &   424.3   &   1.95    &   P$-$L / this paper  &   $-$5.34 &   I   \\
NZ Cen  &   M   &   382 &   2.22    &   P$-$L / this paper  &   $-$5.17 &   I   \\
R Cam   &   M   &   270.22  &   0.84    &   P$-$L / this paper  &   $-$4.81 &   E   \\
V762 Sgr    &   M   &   444 &   1.50    &   P$-$L / this paper  &   $-$5.42 &   I   \\
V407 Sco    &   M   &   396 &   2.11    &   P$-$L / this paper  &   $-$5.26 &   I   \\
ER Aql  &   M   &   337.6   &   1.04    &   P$-$L / this paper  &   $-$5.10 &   I   \\
V441 Cas    &   M   &   175.6   &   1.40    &   P$-$L / this paper  &   $-$4.17 &   \textbf{$\underline{E}$}    \\
EO Cas  &   M   &   455 &   1.83    &   P$-$L / this paper  &   $-$5.47 &   I   \\
IW Cas  &   M   &   396.38  &   1.34    &   P$-$L / this paper  &   $-$5.19 &   $-$ \\
CSS2 10 &   M   &   250 &   3.78    &   P$-$L / this paper  &   $-$4.69 &   $-$ \\
\hline
AU Car  &   M   &   332 &   2.54    &   P$-$L / this paper  &   $-$5.05 &   $-$ \\
RX Car  &   M   &   332.8   &   2.75    &   P$-$L / this paper  &   $-$5.04 &   $-$ \\
\hline
W And   &   M   &   395.93  &   0.38    &   P$-$L / this paper  &   $-$5.27 &   I   \\
RR Mon  &   M   &   394.7   &   1.28    &   P$-$L / this paper  &   $-$5.24 &   I   \\
SU Pup  &   M   &   339.8   &   1.26    &   P$-$L / this paper  &   $-$5.01 &   I   \\
UU Vel  &   M   &   408.9   &   1.31    &   P$-$L / this paper  &   $-$5.32 &   I   \\
RT Sco  &   M   &   449.04  &   0.45    &   P$-$L / this paper  &   $-$5.44 &   I   \\
V865 Aql    &   M   &   364.8   &   0.62    &   P$-$L / this paper  &   $-$5.18 &   $-$ \\
RW And  &   M   &   430.3   &   0.86    &   P$-$L / this paper  &   $-$5.36 &   $-$ \\
AK Mon  &   M   &   328.6   &   2.12    &   P$-$L / this paper  &   $-$5.02 &   $-$ \\
V745 Sgr    &   M   &   380.2   &   0.99    &   P$-$L / this paper  &   $-$5.23 &   $-$ \\
S Cyg   &   M   &   322.93  &   1.94    &   P$-$L / this paper  &   $-$5.06 &   I   \\
FF Cyg  &   M   &   323.82  &   0.87    &   P$-$L / this paper  &   $-$5.07 &   I   \\
\hline
FG Per  &   M   &   340.3   &   3.16    &   P$-$L / this paper  &   $-$5.10 &   $-$ \\
\hline
TT Cen  &   M   &   462 &   1.39    &   P$-$L / this paper  &   $-$5.48 &   $-$ \\
ST Sgr  &   M   &   395.12  &   0.76    &   P$-$L / this paper  &   $-$5.24 &   I   \\
LX Cyg  &   M   &   465.3   &   1.53    &   P$-$L / this paper  &   $-$5.51 &   I   \\
\hline
VX Aql  &   M   &   604 &   1.99    &   P$-$L / this paper  &   $-$5.87 &   I   \\
    &       &   \citet{zijlstra}    &       &       &       &       \\
\hline
V418 Cas    &   M   &   480 &   2.24    &   P$-$L / this paper  &   $-$5.50 &   $-$ \\
\hline              \hline
\end{tabular}
\end{table*}

In a previous paper of this series, hereafter referred to as
"Paper I" \citep{guandalini}, we analyzed a sample of C stars
reconstructing their SEDs up to far infrared, on the basis of
space-borne infrared observations from the ISO and MSX missions.
We found evidence for a relatively large average C-star
luminosity, thus suggesting that the so-called "C-star luminosity
problem" \citep{cohen} might not be real, being simply an effect
of poor estimates of the luminosity, due to insufficient knowledge
of the mid-infrared emission. We also reviewed the available mass
loss rates and showed their correlation with infrared colors.

We want now to extend that analysis, considering those
thermally-pulsing AGB stars where the enhancement of C (and
s-elements) is more moderate than in C stars: it is the case of MS
and S giants \citep{busso92,busso95}. In Sect. \ref{sect2} we
present the sample stars, and we discuss the choices made in
selecting and organizing them in sub-samples, according to the
quality of the available data. In Sect. \ref{sect3} we present the
IR colors and derive the bolometric corrections, based on a set of
sources whose magnitude can be estimated safely through the
integral of detailed SEDs. We also use these corrections for
inferring the critical parameters (absolute Magnitudes or
distances) of sources for which either i) we have incomplete IR
coverage but reliable distance estimates; or ii) Period-Luminosity
relations yield the Luminosity, and the distance needs to be
inferred from the distance modulus. (For the sake of clarity, the
adopted Period-Luminosity relations are discussed in Appendix
\ref{app1}). Once the absolute Magnitudes are known, in Sect.
\ref{sect4} we can analyze HR diagrams and luminosity functions
and on this base we also attempt a rough estimate of photometric
parallaxes for Mira variables with no other available data on
Luminosities. Then, in Sect. \ref{sect5} some preliminary
conclusions are derived, while we postpone to a forthcoming
dedicated work the analysis of stellar winds.

\section{The Sample of S Stars \label{sect2}}

We started from the extended lists by Stephenson
\citep{steph1,steph2}, containing O-rich evolved red giants with
known or suspected "S star-like" chemical peculiarities. As
mentioned, these last anomalies can be due either to in-situ
dredge-up of newly produced elements (intrinsic-S stars) or to
mass-transfer episodes in a binary system (extrinsic-S stars).
From those catalogues we selected a total of 613 sources for which
measurements in the near-IR are available \citep[from 2MASS,
see][]{cutri}. Another required property for selection was the
existence of mid-infrared photometry (although with varied detail
and spectral coverage). The chosen sources fall into various
categories, depending on the extension and quality of the
information we could collect on their IR colors, distance,
variability type, period, luminosity. As a result, we can organize
the stars selected in the following four sub-samples:

\begin{itemize}

\item{\textbf{Sub-sample $A$}. This contains 21 sources for which
ISO-SWS measurements exist up to long wavelengths (40 $-$ 45
$\mu$m), so that a detailed Spectral Energy Distribution (SED) can
be obtained. The mid-IR SEDs can then be integrated together with
near infrared photometric points, in order to infer a very
reliable (apparent) bolometric magnitude, by means of the
relations of fundamental photometry \citep[][see also Paper
I]{glass}. In this way we have a means for computing bolometric
corrections and look for their correlations with available
parameters. Much like for C stars (Paper I), these bolometric
corrections are correlated to near-to-mid infrared color indexes.
About 50\% of the stars in sub-sample $A$ have also astrometric
estimates of the distance from the revised Hipparcos catalogue
\citep{vleu2007,vleu2007b}; for some others we could infer a
distance from the variability period. In all such cases, also the
{\it absolute} Magnitudes can be obtained: the stars with these
characteristics are therefore the fundamental bricks on which we
build the rest of our work. In particular, bolometric corrections
offer us the tools for deriving bolometric Magnitudes for AGB
stars outside this main sub-sample $A$, whenever observations of
at least one near-to-mid color index exist.}

\item{\textbf{Sub-sample $B$}. Here we put all sources (21) for
which a reliable estimate of the distance is available \citep[e.g.
from Hipparcos, especially after the recent revision
by][]{vleu2007,vleu2007b}, but a detailed mid-infrared SED is not
present. Although the stars of sub-sample $B$ have only sparse
mid-infrared photometry (from MSX, IRAS-LRS or ground-based
measurements), thanks to the bolometric corrections discussed
above absolute bolometric Magnitudes can be computed for all of
them.}

\item{\textbf{Sub-sample $C$}. This class contains 41 Mira-type
stars, for which near-to-mid IR colors could be derived from
either MSX or IRAS measurements. Their Mira variability, of known
period, allows for the application of Period-Luminosity relations,
yielding the absolute luminosities \citep[see in
particular][]{whitelock08}. Some comments on the procedure used in
this case to homogenize Period-Luminosity relations are presented
in Appendix \ref{app1}. By also applying the bolometric
corrections computed from sub-sample $A$ to their photometric
data, the apparent bolometric magnitudes can be derived, so that
an estimate for the distance can be inferred from the distance
modulus.}

\item{\textbf{Sub-sample $D$}. Here we collect all the remaining
sources ($\sim$500). One or a few measurements at mid-infrared
wavelengths exist also for this sub-sample; these data can be
coupled to 2MASS photometry, giving near-to-mid infrared colors on
which to apply the bolometric corrections. In this way, we can at
least derive reliable bolometric (apparent) magnitudes. As the
distance is in general not known from measurements (unlike for
sub-sample $B$), and neither it can inferred from any known period
(unlike for sub-sample $C$), a precise value for the absolute
Magnitude cannot be given at this stage and will have to wait for
estimates of the distance. However, we shall see that, at least
for known Mira variables, fiducial intervals for the distance can
be derived from the HR diagram and the luminosity functions, in
the form of a rough photometric parallax}.

\end{itemize}

The first three samples, together with the input data we shall
analyze for them, and the resulting estimates for their
fundamental parameters are included in this note as Tables
\ref{table:1} to \ref{table:9}. The tables of the last sample are
placed in the "online material" and are available electronically.

In general, we perform the same kind of analysis already presented
in Paper I for C-rich AGB stars, computing the bolometric
corrections and bolometric magnitudes, the absolute Magnitudes
when possible, and expressing the color data in the photometric
system described in Paper I and in \citet{busso07}, hereafter
Paper II. In order to adopt this photometric system, we need to
perform a re-binning of space-borne spectroscopic data, by
convolving them with the response of the filters. We then discuss
luminosities and luminosity functions for understanding the
evolutionary relations between S stars and C stars along the
thermally-pulsing AGB.

\section{Infrared Colors and Bolometric Corrections \label{sect3}}

\begin{figure*}[t!!]
\centering \resizebox{\textwidth}{!}
{\includegraphics[height=10cm]{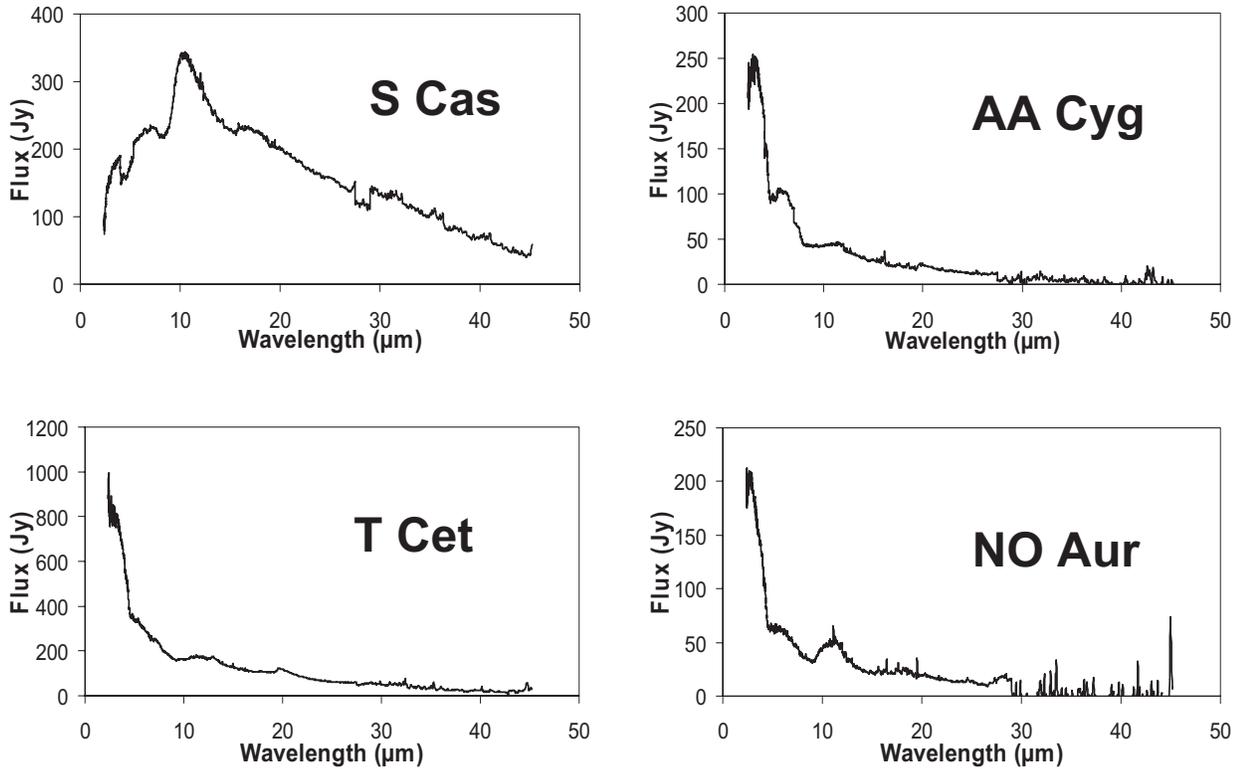}} \caption{Examples of
ISO-SWS spectra of AGB Stars. Upper left: a S-type Mira Variable
(\object{S Cas}). Upper right: a S-type Semiregular source
(\object{AA Cyg}). Lower left and right: two MS stars.
\label{fig1}}
\end{figure*}

We have made a preliminary control that the choice of near and mid
IR wavelengths is sufficient for our scope. In fact, one might a
priori believe that the inclusion of optical bands could increase
the accuracy of bolometric corrections (despite the large
uncertainties related to the photospheric variability). However, for
all the stars of our samples for which we could find archived visual
magnitudes, the optical-to-near-IR colors (e.g. V-J) are always in
excess of 3mag. Most V-J colors lay in the range 4-7 and a few in
the range 7-10mag, implying flux ratios $f_J/f_V$ larger than at
least a factor of 10 and more often a factor of 100. There would be
therefore no significant change in the results by introducing
optical photometric data. S stars are certainly bluer than C stars
(only a couple of sources are dominated by mid-IR, as is instead
common for C-rich Miras, like CW Leo); but even when the infrared
excess by dust is minimal and the photospheres are optically bright,
all AGB stars, including S stars, remain mainly IR objects. The
bluest of our sources have maximum emission in the H band.

As discussed in the previous section, we then start from the sources
of sub-sample $A$ in order to derive the Bolometric Corrections,
then proceed to the other stars in our selection. The relevant
parameters we collect or derive can be divided into three
categories, so we have three tables per each sub-sample. As
mentioned before, we include here those referring to groups $A$,
$B$, and $C$.

Table \ref{table:1} presents the general characteristics of
sub-sample $A$. This includes: i) the names of the sources (from
IRAS, from the Stephenson's compilation \citep{steph1,steph2} and,
when available, from the variable star nomenclature); ii) the
coordinates (with the equinox fixed at year 2000.0) as given by
the ICRS on the SIMBAD database; iii) the spectral types; and iv)
the variability type, according to the General Catalogue of
Variable Stars \citep[GCVS,][]{samus}. Some of the stars we
analyze have uncertain spectral classifications for various
reasons. As this point is not among the main issues we want to
address, in all these cases we indicate the alternative choices
found in the literature, simply separating them by a slash ({/}).
In some cases, our analysis will yield suggestions on the correct
classification.

The ISO-SWS SEDs of a few sources of group $A$ are shown in Fig.
\ref{fig1}, organized according to their spectral classification (MS
or S) and variability type (Mira, Semiregular or Irregular). The
figure shows one of the most evident properties of AGB stars,
present also in C- rich sources: the Mira-type variability is often
associated with a remarkable IR excess and a SED peaked at longer
wavelengths than for Semiregular or Irregular variables. As compared
to C stars, however, the IR excess of S stars is in general much
smaller, as we already mentioned. This is probably related to
intrinsic properties of dust with O-based or C-based composition. A
similar difference exists for molecules (e.g. gaseous component) of
the two species, as seen from the photospheres, as we shall argue
later.

\begin{figure*}[t!!]
\centering \resizebox{\textwidth}{!}
{\includegraphics[width=\textwidth]{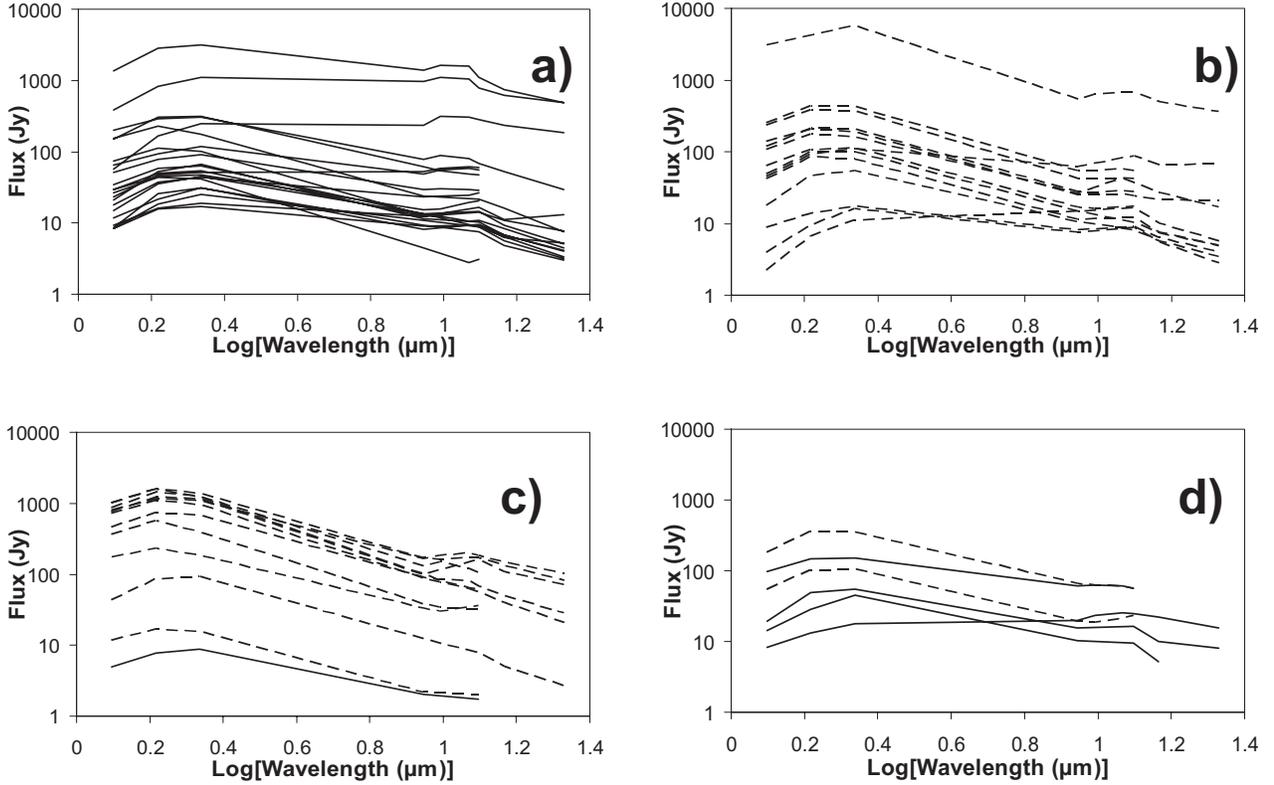}} \caption{Spectral
energy distributions of sample stars after rebinning in the
photometric system of Paper I. a) S Miras. b) S Semiregulars. c)
MS sources. d) SC stars. Continuous lines refer to Miras, dashed
ones to Semiregulars.} \label{fig2}
\end{figure*}

The above very detailed energy distributions are integrated to
estimate the apparent bolometric magnitude, using the relation:
\begin{equation}\label{eq1}
\centering{
 m_{bol} = -2.5 \int_0^\infty f_\nu d\nu + C}
\end{equation}
as discussed by \citet{glass}, who also gives $C$ = $-$18.98 in
the case where the total flux is expressed in W m$^{-2}$. This
relation also provides the zero-magnitude flux. From Equation
\ref{eq1} one obtains:
\begin{equation}\label{eq2}
\centering{
 f_0 = \int_0^\infty f_\nu d\nu  = 1.94\times10^{-7} {\rm W m^{-2}} =
 1.94\times10^{19} {\rm Jy Hz}}
\end{equation}
Following the procedure described in Papers I and II, we again
operate a re-binning of the ISO spectra to derive estimates for
the fluxes in a number of broader-band infrared filters (with 10\%
bandwidth), in the range between 8 and 14 $\mu$m (centered at 8.8,
9.8, 11.7 and 12.5 $\mu$m). (The fluxes in such filters are
usually indicated as the equivalent wavelengths in normal
brackets; the magnitudes use a similar notation but with squared
brackets). These filters are of rather common use in ground-based
infrared cameras; our choice is aimed at making the space-borne
measurements more easily comparable with present and future
photometric observations and to infer for these last suitable
bolometric corrections. At long wavelengths ($\lambda >$ 14
$\mu$m) the re-binning is made using the response curve of the MSX
filters $D$ (14.6 $\mu$m) and $E$ (21.3 $\mu$m). Zero-magnitude
fluxes (in Jy) for our chosen photometric system are: 52.23 (8.8
$\mu$m), 42.07 (9.8 $\mu$m), 29.55 (11.7 $\mu$m), 25.88 (12.5
$\mu$m), 20.25 (14.6 $\mu$m), 8.91 (21.3 $\mu$m). For near-IR,
2MASS calibrations for the J, H and K$_s$ filters are given in
\citet{cohen03}.[Hereafter we shall adopt for the K "short" K$_s$
filter the simple notation K].

Figure \ref{fig2} gives some examples of the SEDs computed in the
photometric system thus defined, for the various types of AGB
stars discussed in this note. The same data are listed in Tables
\ref{table:2}, \ref{table:5} and \ref{table:8}, together with
near-infrared fluxed obtained from the 2MASS catalogue.

As a consistency check, we repeated the calculations of Equation
\ref{eq1} using the apparent bolometric magnitudes previously
derived from ISO spectra, and making the integral over the
low-resolution SEDs of Fig. \ref{fig2}. By adopting the
zero-magnitude fluxes by \citet{bessell}, we obtain for the
integration constant an average value $C$ = $-$19.0. The fact that
this estimate be very close to the real $C$ value given above
certifies that the our re-binning is done properly, and does not
alter in any significant way the integrals of Equation \ref{eq1},
hence the magnitudes of our stars.

\begin{figure*}[t!!]
\centering \resizebox{\textwidth}{!}
{\includegraphics[height=5cm]{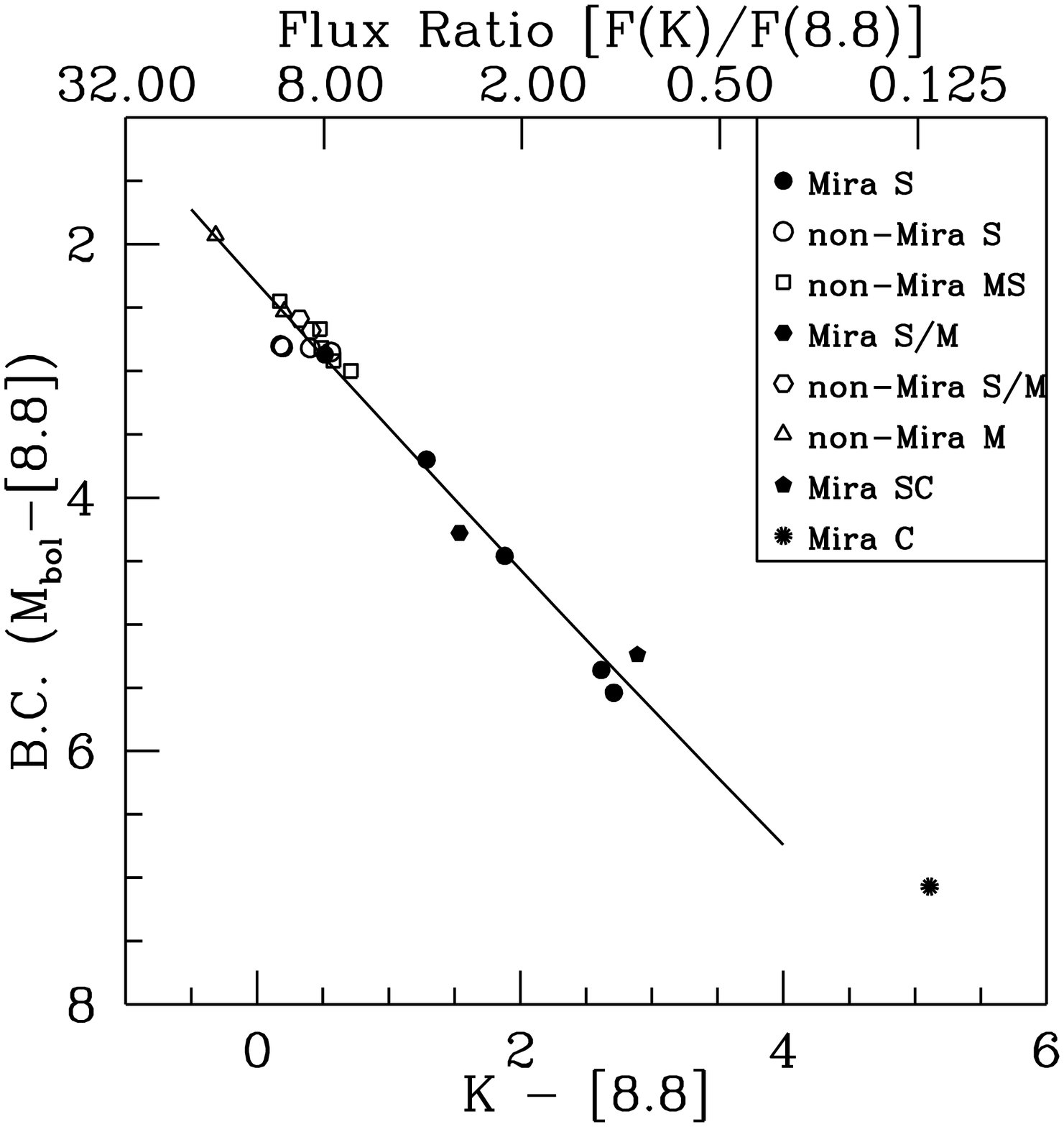}
\includegraphics[height=5cm]{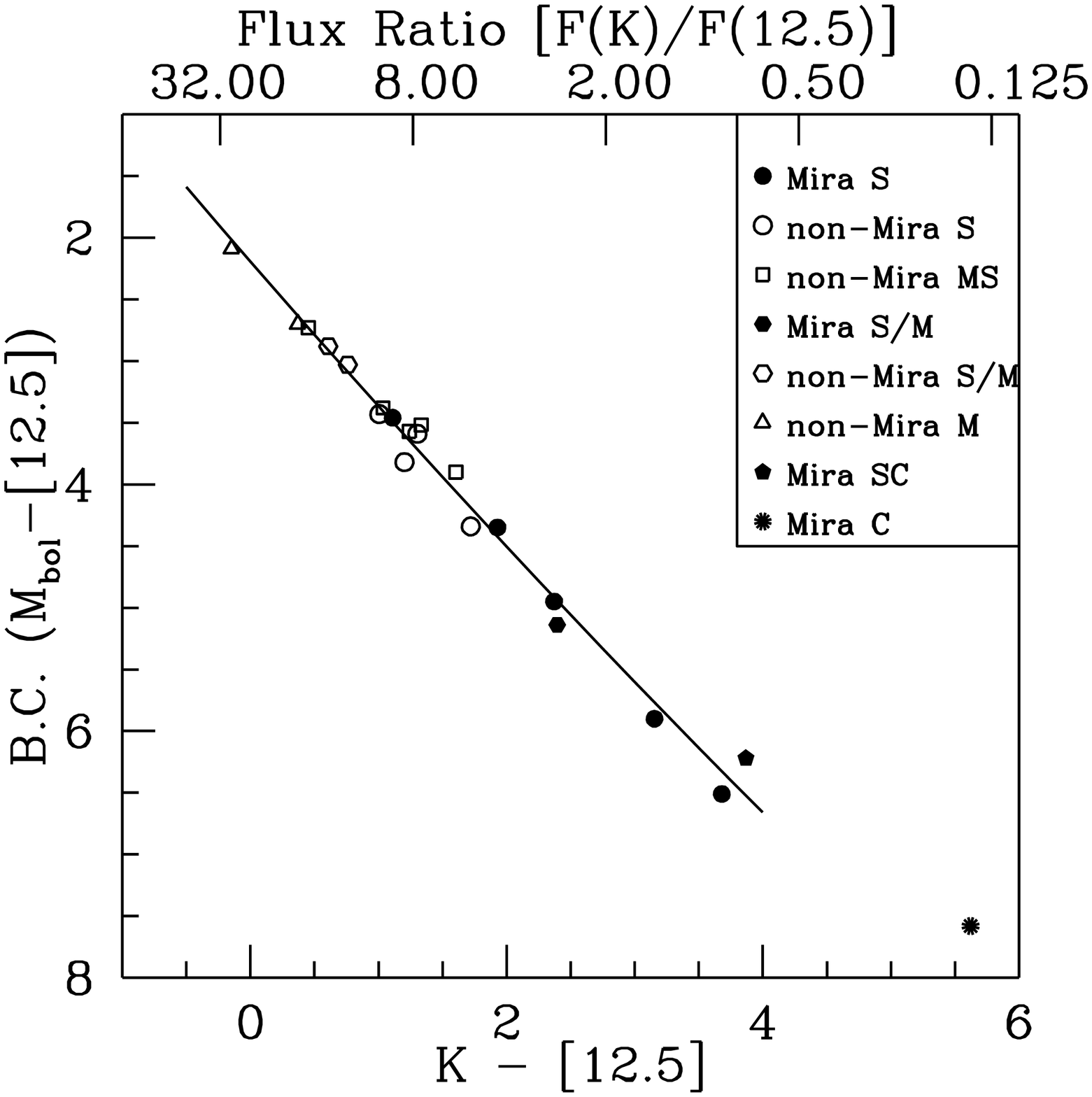}}
\caption{Two examples of bolometric corrections, suitable for MS-S
Stars with infrared photometry available. Correlation coefficients
for the two relations are $R^2 = 0.98$ (left) and $R^2 = 0.99$
(right). Here and elsewhere "K" is a compact notation for the
2MASS filter K$_s$ (K-short).} \label{fig3}
\end{figure*}

Once the apparent bolometric magnitudes are known, for those
sources for which a reliable estimate of the distance exists we
can derive $M_{bol}$. This estimate can be obtained also for those
Mira variables of sub-sample $A$ for which an astrometric
measurement of the distance is not available, but the variability
period is known. Indeed, the Period-Luminosity relations of Mira
stars have become sufficiently reliable that we can use them to
infer $M_{bol}$ and then obtain the distance from the distance
modulus (see Appendix \ref{app1}). The limit of this technique
lays in the fact that, not having a proper description of the
interstellar extinction, we are forced to exclude the extinction
correction. The error introduced on infrared colors, however, is
not large, and the related uncertainty remains within the (broad)
observational errors of typical infrared photometry (see Paper I
for a discussion).

Table \ref{table:3} presents the absolute bolometric Magnitudes
derived for sub-sample $A$, with the indication of the method
used, i.e. Period-Luminosity relations (labeled "this paper") or
distances in the literature (labeled after the reference adopted,
either "Hipparcos", see \citet{vleu2007}, or in a single case
"Groenewegen", taken from \citet{groenewegen02b}). The values
adopted for either the period or the distance are also indicated.

We can then correlate the newly found bolometric Magnitudes with
color indexes in the infrared, in order to infer bolometric
corrections to be applied to the other groups of stars in our
sample. In Fig. \ref{fig3} we present two of the most significant
such correlations. They are shown as corrections to a mid-infrared
magnitude ([8.8] or [12.5]) as functions of near-to-mid infrared
color indexes (K-[8.8] or K-[12.5] respectively). Least-square
fits, shown in the panels of the figure, correspond to:
\begin{equation} \label{eq3}
B.C. = -0.0118 \cdot (K-[8.8])^2 + 1.1552 \cdot (K-[8.8]) + 2.3061
\end{equation}
and to
\begin{equation} \label{eq4}
B.C. = -0.0195 \cdot (K-[12.5])^2 +1.1949 \cdot (K-[12.5]) +
2.1918
\end{equation}
with correlation coefficients of $R = 0.990$ and $R = 0.993$,
respectively.

As shown in the figures, both relations are quite tight and offer
a reliable way to estimate bolometric Magnitudes for stars for
which we have only partial photometric coverage. In the two plots
of Fig. \ref{fig3}, there is only one point which is significantly
discrepant and does not follow the average correlation. Quite
significantly, this point corresponds to \emph{II Lup}, a star
that, although listed in the Stephenson's compilation of S stars,
is instead a carbon-rich Mira variable and must therefore be
excluded in discussing the photometric properties of O-rich AGB
stars.

At this point we need to underline a relevant characteristic of
the data displayed in Fig. \ref{fig3}. The corrections there used
to infer the bolometric emission are applied to mid-infrared
magnitudes, i.e. to data that, for our S stars with moderate IR
excess, are essentially constant. This constancy of the mid-IR
emission is important for estimating better bolometric corrections
and is not shared by very red objects, e.g., by the carbon-rich
Miras, where also the emission from the circumstellar envelope
varies significantly in time (see Paper II). Anyhow, for our
sample stars in Figure \ref{fig3} we expect roughly constant
magnitudes at 8.8 or 12.5 $\mu$m, while the K(short) estimates
from 2MASS, being single-epoch measurements, are affected by the
photospheric variability (that can be up to 1 - 1.5 mag in
near-IR). Different stars were certainly (randomly) sampled by
2MASS in different epochs of their light curves, but despite this,
all the data fit a single regression line quite well. This can be
so only if, during a cycle, the representative points of our stars
move along the regression line itself. One might wonder why. The
variability of AGB stars is not necessarily related only to a
transfer of energy from one wavelength range to another (due to
the star becoming cooler and more extended or hotter and more
compact). There might be actually a global variability of the
bolometric magnitude itself, because we can see the addition, with
different percent weight, of energy from finite-amplitude
mechanical oscillations and shock waves affecting the envelope.
The color dependence of these emissions is unknown to us, but we
know experimentally that it does not extend to mid-IR for
moderately red objects. If the chosen mid-IR magnitudes remain
essentially constant, then bolometric corrections variable in time
will be needed to compensate for the near-IR (and for the possible
bolometric) variability. When the star is brighter it will require
a larger correction, when it is dimmer the correction will be
smaller, but as long as the regression line remains valid, the
given formula for the B.C. should continue to hold. This would not
be true if we had chosen, in the ordinate, a near-IR magnitude,
still affected by the photospheric variability, having an unknown
color dependence.

\begin{figure*}[t!!]
\centering
{\includegraphics[height=9cm]{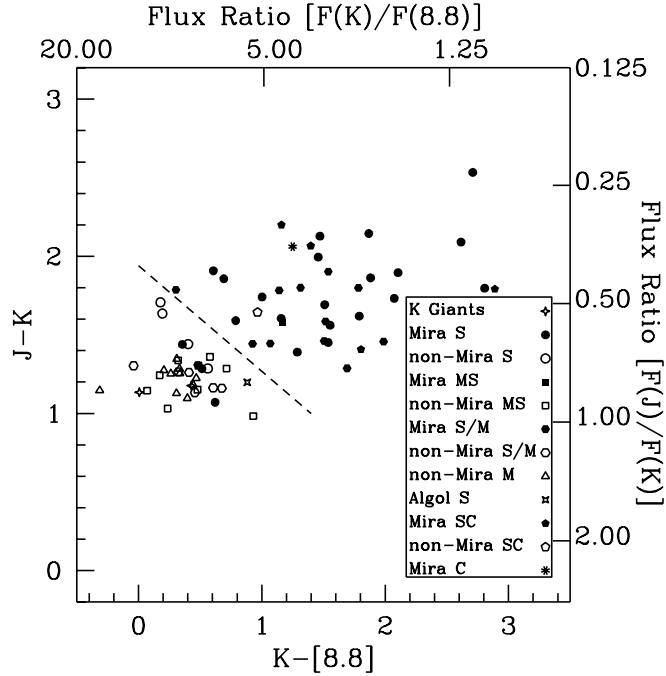}} \caption{A color-color
diagram of the sources in samples A to C. All of them belong to
the Stephenson's list, but some are misclassified: indeed, we find
a few K and M giants, and a couple of C stars. Mira variables are
redder than other sources in both colors and generally well
separated from the rest by IR color criteria.} \label{fig4}
\end{figure*}

It remains however true the bolometric magnitudes given here,
although possibly rather well estimated thanks to the use of
mid-IR data, might be intrinsically variable. We have estimated
how much the rather common variability by about 1 mag in K$_s$
would affect the global flux, and this can reach up to 25\% of the
total. When estimating bolometric Magnitudes, throughout this
paper, we apply both the corrections of Fig. \ref{fig3} whenever
possible, and then take the average of the absolute Magnitudes
thus obtained (the values to be averaged are in any case very
close to each other).

We then applied the above corrections to the IR colors of
sub-sample $B$, thus deriving also for them an estimate of the
bolometric Magnitude. The input data and the derived parameters
for this sample are shown in Tables \ref{table:4} to
\ref{table:6}.

A somewhat inverse procedure is applied to the sources of sample
$C$, where Luminosities can be inferred from Period-Luminosity
relations. Here a comparison with the apparent bolometric
magnitude, derived from bolometric corrections, yields the
distance of the object.  The recent updates in Period-Luminosity
relations for AGB stars, which now allow for this possibility, are
discussed in Appendix \ref{app1}.

The input data and the derived parameters for sample $C$ are
illustrated in Tables \ref{table:7} to \ref{table:9}. The whole
(more sparse) information we could compile for the bigger but so
far incomplete sub-sample $D$ is included instead in Tables
\ref{table:11} and \ref{table:12}, which we publish in electronic
form due to their dimensions. For the content of Table
\ref{table:10} see next section.

\section{HR Diagrams and Luminosity Functions \label{sect4}}

Using the IR fluxes collected in the previous section, and the
information acquired on the absolute luminosities (see also
Appendix \ref{app1}), we can now study the photometric properties
of S stars and their relatives, with the aim of understanding how
they are linked to the evolutionary status of the sources and to
their chemical properties.

Figure \ref{fig4} illustrates the J-K vs K-[8.8] color-color
diagram of the sources in our sample. The variability types appear
rather well discriminated from IR colors. With only one exception
(an SC star believed to be of Semiregular variability, but that we
suspect is instead a misclassified Mira), all Semiregular and
Irregular variables lay in the lower-left part of diagram, and
remain separated by Miras by a dividing gap (the dashed line is
just an eye's guide to illustrate this). Similarly, only very few
objects classified as Miras fall at the left of the guiding line,
all of them with short periods, while the vast majority is, in
both colors, redder than Semiregulars and Irregulars. This is not
surprising, in view of the fact that Miras are more efficient
mass-losers and, as we argue below, are on average more evolved.
We actually interpret the emerging evidence (from both the present
sample and those of Papers I and II) as a suggestion that
efficient radiation pressure on dust grains, powering fast mass
lost, starts when the star has reached the Mira-like variability
at periods in excess of about $200-230$ days. In any case, the use
of IR colors for discriminating different sources is interesting
because the separation of the types is quite good, to the point
that the IR colors might be used for guessing the variability
type. (For example, if one picks up an AGB star of unknown
variability, with J-K and K-[8.8] colors near 2, then we can
reliably assume that it must be a Mira star).

\begin{figure*}[t!!]
\centering
{\includegraphics[width=8cm,angle=-90]{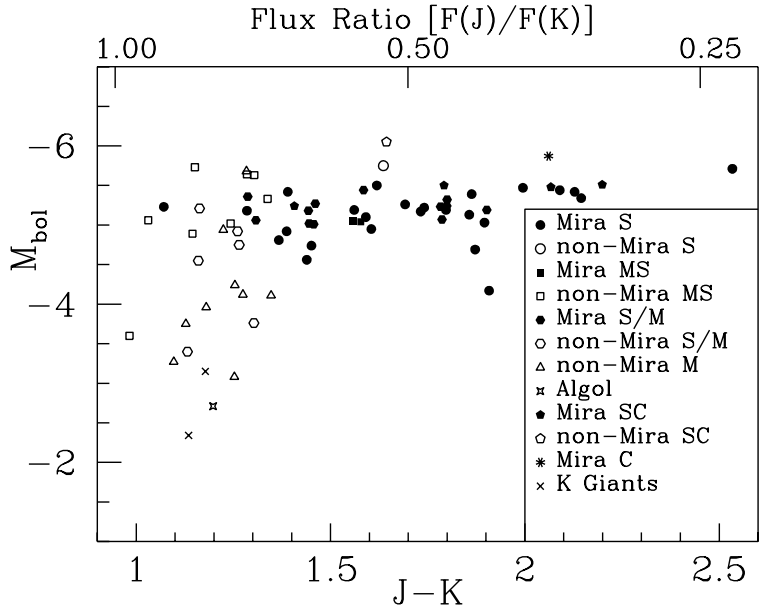}} \caption{An
example of an absolute HR diagram, built using a near IR color
index as abscissa.} \label{fig5}
\end{figure*}

\begin{figure*}[t!!]
\centering
{\includegraphics[width=8cm,angle=-90]{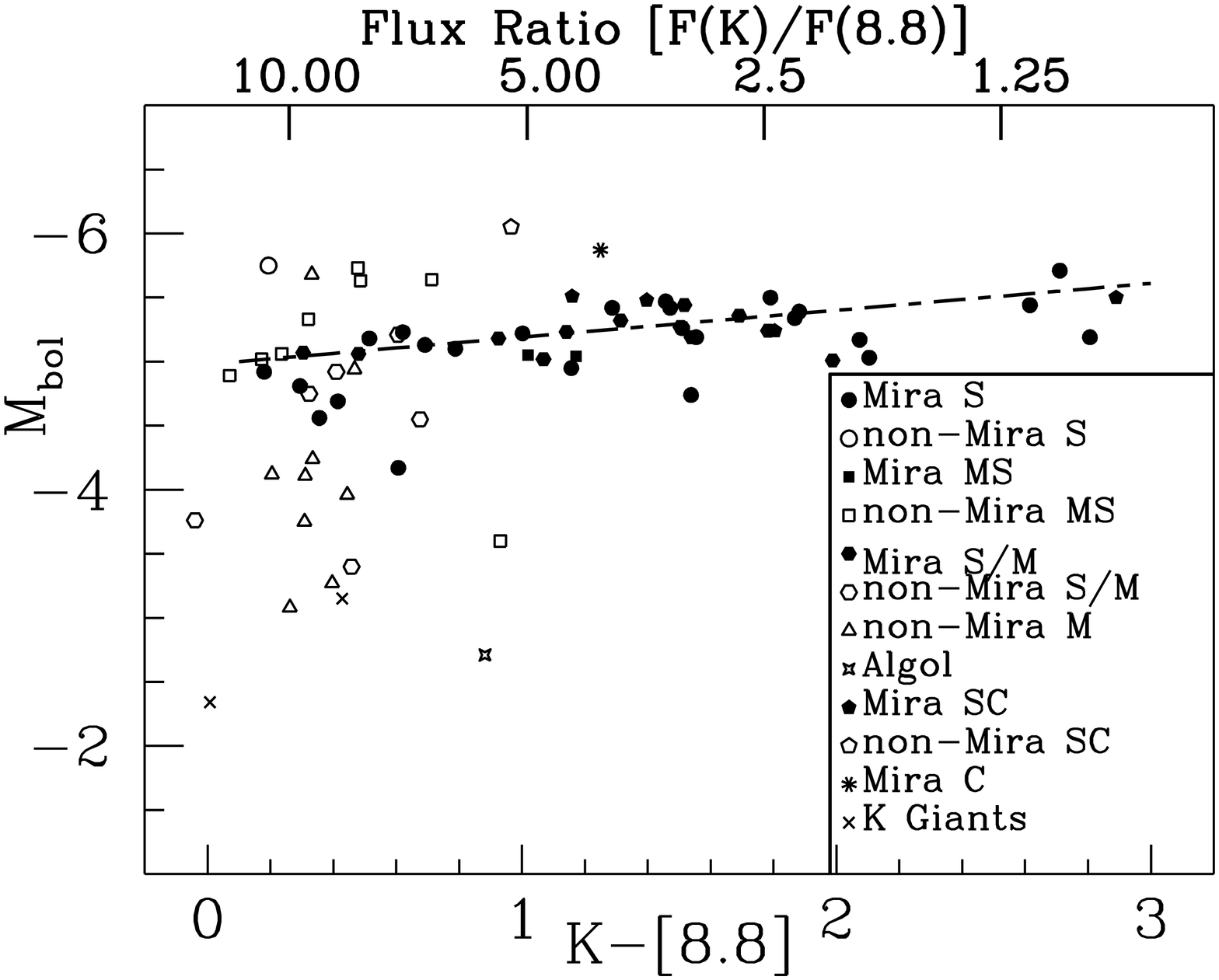}} \caption{An
example of an absolute HR diagram, built using a near-to-mid IR
color index as abscissa.} \label{fig6}
\end{figure*}

Figures \ref{fig5} and \ref{fig6} show two examples of the absolute
H-R diagrams derived from our sample stars (the figures refer to
both intrinsic and extrinsic sources). They are presented as a
function of J-K and K-[8.8] colors, which can be considered as
monitors either of the photosphere alone (J-K) or of the inner
circumstellar environment (K-[8.8]). The link between fluxes in Jy
and magnitudes in the various filters, hence to colors, is
established by i) the distance correction (no extinction is
assumed): and ii) the calibration of zero-magnitude fluxes given
above.

The HR diagrams contain a large number of useful pieces of
information on S Star morphology and physics. We identify here a
few such issues.
\begin{itemize}

\item{Absolute Magnitudes of Semiregular and Irregular variables
stay on an almost vertical sequence, close to the region covered
by AGB stellar models, albeit with remarkable scatter. Below this
sequence, at the bottom of the plot, we find sources whose
luminosities suggest that they are in an evolutionary stage
preceding the thermally-pulsing AGB. These must necessarily be
extrinsic S or MS stars; indeed, in most cases their extrinsic
nature is already well established. For those sources (five) in
this group for which no information on binarity is available, we
suggest here, as a result of our luminosity calibration, that they
are extrinsic and not yet on the TP-AGB (see the relative
indication in bold and underlined in Tables \ref{table:3},
\ref{table:6}, \ref{table:9}). This suggestion will now need
independent confirmations, which would also indirectly verify (or
deny!) the validity of our bolometric corrections. If we are
right, then extrinsic S stars should be the O-rich equivalent of
the carbon rich R stars; Fig. \ref{fig5} supports this guess in
showing that lower-luminosity S stars are also warmer than the
others.}

\item{S-star infrared colors are different from those of C stars,
being on average much bluer. Infrared excesses are less extreme,
although mid-infrared (especially the [8.8] filter) remains the
crucial wavelength range for understanding their physics: in
particular we underline that, also for S stars, the best relations
determining bolometric corrections are obtained by making use of
mid-infrared colors.}

\begin{table*}
\caption{Sample D $-$ Distance Estimates.}             
\label{table:10}      
\centering                          
\begin{tabular}{c c c c}        
\hline      \hline
Source  &   Apparent bol. &   Distance    &   Min. $-$ Max. Distance  \\
name    &   magnitudes    &   (kpc)   &   (kpc)   \\
\hline      \hline
NX Per  &   7.61    &   3.57    &   2.97 $-$ 4.30   \\
NU Pup  &   8.58    &   5.56    &   4.63 $-$ 6.69   \\
DK CMa  &   6.06    &   1.75    &   1.45 $-$ 2.10   \\
EW Pup  &   7.04    &   2.74    &   2.28 $-$ 3.29   \\
WY Pyx  &   5.01    &   1.08    &   0.90 $-$ 1.29   \\
V2434 Oph   &   4.95    &   1.05    &   0.87 $-$ 1.26   \\
V342 Ser    &   6.55    &   2.18    &   1.82 $-$ 2.63   \\
V471 Sct    &   5.85    &   1.59    &   1.32 $-$ 1.91   \\
V427 Sct    &   6.88    &   2.54    &   2.11 $-$ 3.06   \\
V1959 Cyg   &   6.47    &   2.11    &   1.76 $-$ 2.54   \\
\hline
PR Nor  &   4.51    &   0.85    &   0.71 $-$ 1.03   \\
PZ Vul  &   8.43    &   5.20    &   4.33 $-$ 6.25   \\
GY Lac  &   7.31    &   3.11    &   2.58 $-$ 3.74   \\
V928 Cas    &   8.22    &   4.73    &   3.93 $-$ 5.69   \\
V508 Aur    &   8.28    &   4.86    &   4.04 $-$ 5.84   \\
V1992 Cyg   &   7.90    &   4.07    &   3.38 $-$ 4.89   \\
V1850 Cyg   &   8.56    &   5.53    &   4.60 $-$ 6.65   \\
V1242 Cyg   &   8.12    &   4.50    &   3.74 $-$ 5.41   \\
\hline      \hline
\end{tabular}
\end{table*}

\item{The un-reddened MS-S sources distributed along the vertical
branch in Figs. \ref{fig5} and \ref{fig6} are more compatible with
the evolutionary AGB tracks than for similarly-positioned C stars.
Red-shifts of the track, requiring temperature corrections (and
indicating inadequacy in molecular opacities of the envelope),
although present, are largely reduced (by 0.5$-$1 mag.) as
compared to C stars (see Paper II).}

\item{Absolute Magnitudes of Mira variables are distributed over a
relatively well defined distribution. They show a linear trend,
indicating higher luminosities for redder colors and the
correlation is clear, although the slope is small. In Fig.
\ref{fig6} we have added a least square curve based only on
bona-fide, intrinsic S Miras. The relation is:
\begin{equation} \label{eq5}
M_{bol} = -0.2104 \cdot (K-[8.8]) -4.9765
\end{equation}
with a correlation coefficient $R = 0.78$. Of course one expects
different luminosities for sources of different colors
(Period-Luminosity relations are such that the LPVs of longer
period are also redder). It is however interesting to notice that,
despite the infrared excesses increasingly separate the observed
points from the AGB model tracks, we can nevertheless determine a
systematic trend in the magnitude (now dominated by the
circumstellar envelope). Moreover, it's remarkable that the
correlation is sufficiently tight not to be confused by the
observational errors.

If one computes a global average of the Magnitudes of Miras, this
turns out to be $-5.15\pm0.4$. Once stars have reached this
luminosity range and are in the Mira variability stripe, further
remarkable increases of luminosity seem to be no longer possible.
This limiting range of luminosities (less than 1 mag. wide) is
close to a similar one that can be inferred for C-rich Miras in
the data of Paper I. Since the range is relatively small, Mira
variables appear to have, on average, similar Magnitudes,
independent of the chemical composition of the atmosphere.}

\item{We can confirm for S stars what was already said for
C-stars: the Mira variability type tends to occur primarily at the
end of the evolution. It was argued in the past that Semiregulars
might be AGB stars in the low-luminosity post-flash dip
\citep{kersch}, so that a repeated transition between the two main
variability types was expected. We cannot exclude this in general,
but the high statistical relevance of Semiregular variables
(almost as abundant as Miras in our global database, including
sources of group $D$) cannot be explained by the post-flash
phases, which, for advanced thermal pulses, occupy at most 25\% of
the TP-AGB duration (see also Fig. \ref{fig8}).}

\item{The limited spread of Mira Magnitudes and the sufficiently
good correlation with IR colors offer a tool for obtaining a
first-order approximation to the luminosity and distance of Mira
stars for which we do not have information beyond the measured
colors. Indeed, for them one can first derive the apparent
bolometric magnitude from our bolometric corrections; then one can
assume, as a reasonable guess, that bolometric absolute Magnitudes
have an average of $-5.15\pm 0.4$ and a correlation with the
K-[8.8] color like the one shown for Fig. \ref{fig6}. This allows
a determination of the distance from the distance modulus. An
example of the application of this procedure to Mira variables in
sample $D$ is shown in Table \ref{table:10}.}

\item{Stars differently classified along the sequence MS, S, SC
show a different behavior also for what concerns the variability.
Albeit with some scatter, one can notice that the MS classification
is in general (with two possible exceptions) accompanied by the
Semiregular or Irregular variability types, while a much larger
percentage (about 60\%) of S stars shows the Mira-type variability.
Concerning SC stars, of the 5 objects for which we have detailed
information, 4 are of Mira type; from the color-color diagram of
Fig. \ref{fig4} we suspect this should be so also for the fifth
member, for which we suggest that the previous classification as a
Semiregular might be wrong. SC stars are in general red, cool and
luminous, like most S Miras. They might be the final evolutionary
status of stars whose mass is barely sufficient to dredge-up enough
carbon to approach unity in the C/O ratio.}
\end{itemize}

\begin{figure*}[t!!]
\centering
{\includegraphics[width=8cm,angle=-90]{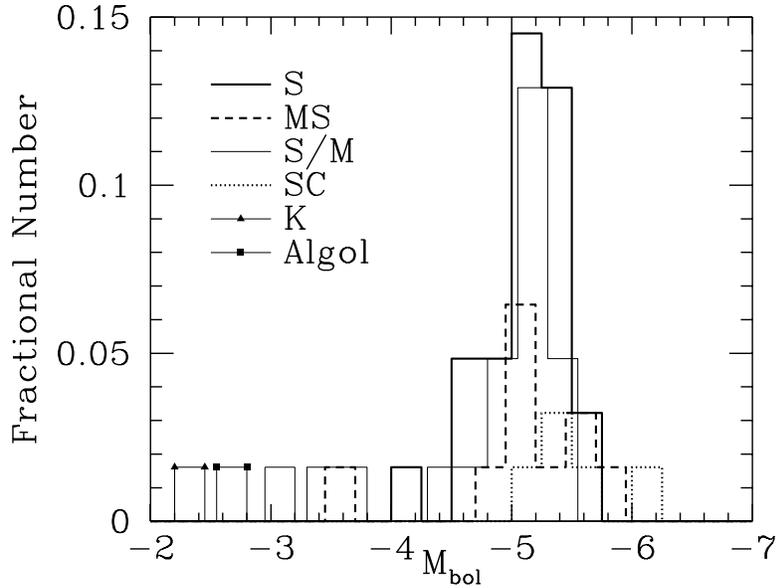}} \caption{The
luminosity function of the stars in our samples $A$ to $C$. Low
luminosity sources are always extrinsic and the S stars define a
very narrow distribution, characterized by a magnitude
$-5.15\pm0.4$.} \label{fig7}
\end{figure*}

The luminosity properties illustrated by the HR diagrams, and in
particular the narrow range over which the magnitudes of MS-S
Miras are distributed, are illustrated by the histogram of Fig.
\ref{fig7}, showing the luminosity function of our sources. The
distribution has a very well defined peak: all MS, S, SC stars lay
in the range $-4$ to $-6$ and Mira S stars occupy a thin slice
across the magnitude values $-4.8 $ to $-5.5$. All the data points
significantly far from this peak refer to extrinsic sources not
belonging to the TP-AGB phase. Many (most) sources of uncertain
classification (so far indicated as S/M) fall in the fiducial
interval of the 'best' S stars. We are pretty sure they are indeed
S stars and probably Miras. In Tables \ref{table:3},
\ref{table:6}, \ref{table:9} and \ref{table:12} we indicate the
possible classification deduced from \citet{vaneck00,yang}. In a
few cases of discordant indications we prefer the physical
analysis by \citet{vaneck00} and we show the other in parenthesis.
For five sources in Tables \ref{table:6} and \ref{table:9}, for
which neither study offers a suggestion, we can infer that they
are extrinsic from their low intrinsic Luminosity.

Like for C stars, and actually even more so, the Luminosities of
Population I S stars are well defined inside a narrow range and are
high enough that normal stellar models based on the Schwarzschild
criterion for convection can adequately explain their formation.
Once again, and as already noticed in Paper I, stellar models using
large convective overshoot to favor dredge-up, and to obtain S-star
and C-star chemical peculiarities at low luminosities \citep{izz07}
appear to be unjustified, at least as far as the AGB evolutionary
stages are concerned. Recently \citet{bona07} suggested that such
models should be preferred for their capability of reproducing
$s$-process abundances in stellar populations.  One has however to
mention that, from a theoretical point of view, the several free
parameters still involved in any modelling probably make any
conclusion provisional, and require observational verifications.
Infrared observations, as presented here, do not confirm the need
for large overshooting, at least for what concerns the ensuing
stellar luminosities.

\section{Conclusions \label{sect5}}

\begin{figure*}[t!!]
\centering \resizebox{\textwidth}{!}
{\includegraphics[height=6cm,angle=-90]{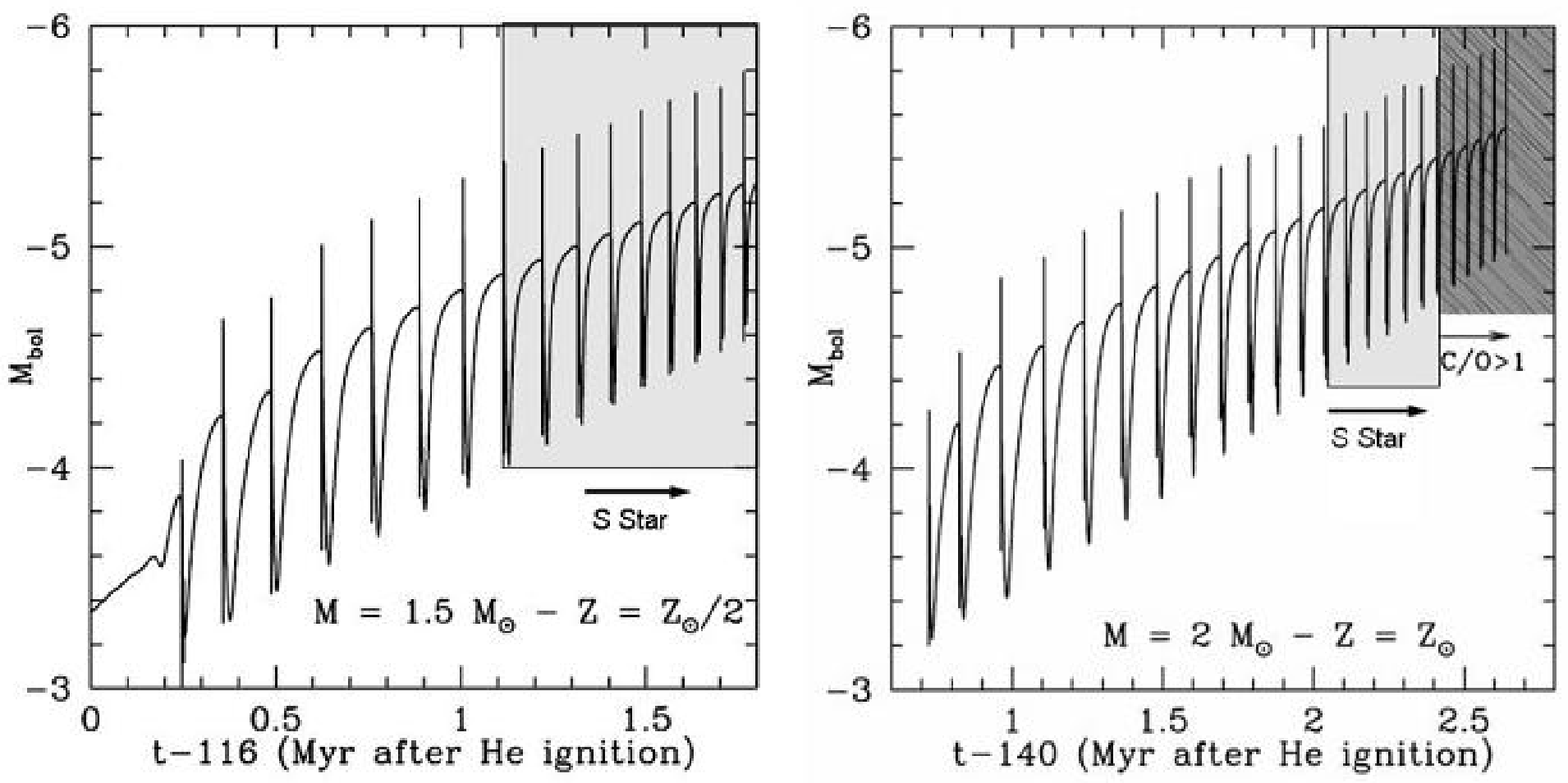}} \caption{The
magnitude variations of two model stars of Population I, during
the thermally pulsing AGB phase, as computed by the FRANEC code.
The abscissa shows the time passed after the ignition of
He-burning in the core, in Million years (once an offset is
deduced, whose value is indicated).  Most sources with features
typical of S or C stars and magnitudes in the range $-5.2 \pm 0.4$
should belong to the mass and metallicity range illustrated here,
but only the star in the right panel does achieve the C/O $>$ 1
condition.} \label{fig8}
\end{figure*}

In this paper we have presented a reanalysis of the properties of
MS-S-SC stars, based on a sample of about 600 sources, whose
infrared fluxes from 1.25 to 21 $\mu$m were measured by the 2MASS,
IRAS, ISO and MSX experiments. A 'best' group of 21 stars (for
which detailed ISO-SWS spectra are available up to long
wavelengths) allowed us to obtain the bolometric Magnitudes from
an effective integral of the spectral energy distribution up to 45
$\mu$m. Correlations with near-to-mid IR colors then allowed us to
infer bolometric corrections suitable to be applied to other
groups of sources, with a less detailed coverage of the energy
distribution. We can thus estimate with sufficient accuracy the
apparent bolometric magnitudes of more than 500 sources. The whole
analysis was performed in the photometric system suggested in
\citet{busso96} and subsequently used in Paper I and Paper II.

Criteria for obtaining the distance have then been discussed, from
the simple use of revised astrometric measurements to a
reformulation of the known Period-Luminosity relations for O-rich
Long Period variables. The results of our analysis suggest that
Mira variables of the S type have on average magnitudes in the
range $-5.15\pm0.4$, showing a well-defined linear correlation
with infrared colors, especially the K-[8.8] one. Low-mass AGB
stars do not appear to proceed beyond the upper limit of the Mira
luminosity range. Inside this range $P-L$ relations have become
rather tight and accurate; they now form a very useful tool for
determining intrinsic stellar parameters. From the $P-L$ relations
and the $M_{bol} - (K-[8.8])$ relation of Equation \ref{eq5} we
now have tools to estimate the absolute Magnitudes, hence the
distances, of Mira S variables for which either the period or the
infrared colors have been determined.

From statistical considerations it is also argued that Mira
variables should occupy mainly the final part of the AGB track, as
the simple intermittency between low-luminosity post flash dips
and high-luminosity H-shell-powered stages is not sufficient to
explain the available numbers of Semiregulars and Miras.

Luminosity functions confirm that intrinsic S stars (especially if
in the Mira class) are distributed over a narrow range around the
above average magnitude, and that this last datum is very close to
the one previously found for C-stars. This is so to the point that
it appears unlikely that a single AGB star can follow the whole
M-MS-S-SC-C sequence, by simply increasing gradually its content
of carbon and heavy elements as time passes, new dredge-up
episodes enrich the envelope and the luminosity increases. Most
probably, small differences in the initial mass and metallicity,
almost indistinguishable once on the AGB, determine the final
chemical fate of a star, which, for increasing initial mass, can
end its life either as a MS-S giant, or a SC giant, or reaching
effectively the C(N) stage. As an example of the effects of small
mass (and metallicity) differences, Fig. \ref{fig8} shows two AGB
luminosity sequences obtained from the FRANEC code by
\citet{busso03}. The model star of the left panel does not reach
the C-star phase, ending as an MS/S star. The evolutionary phases
during which S-type chemical peculiarities are exhibited by the
envelope are shaded in the plot. In the right panel we instead
show the results of a model producing a real C-star: it has a
slightly higher mass. Given the magnitudes we found in this paper
for the sources of types MS and S, and those found in Paper I for
C stars, most of the objects in our sample should belong to the
mass and metallicity range covered by Fig. \ref{fig8}. If we
broadly consider the typical magnitudes of S and C stars as being
in the range $-5.2 \pm$ 0.4, then it is clear that this range
includes both the S-star phase in the first panel of Fig.
\ref{fig8} and the S- and C-star phases in the right panel. It
would be very difficult or impossible to distinguish between the
two cases on the basis of their magnitudes, given the remaining
uncertainties on distances and on possibly variable bolometric
magnitudes.

The above results, and the typical initial mass implied for S and C
stars (1.5 - 2 $M_{\odot}$) might appear rather peculiar, in view of
the fact that about half of Planetary Nebulae are carbon rich. This
cannot be explained by progenitors of around 2 $M_{\odot}$, at least
according to the most common choices for the Initial Mass Function.
However we must remember that the efficiency of dredge up strongly
increases with decreasing metallicity. In halo stars one solar mass
might be sufficient for forming a C star (also because the abundance
of oxygen is very small). Hence the results presented here should be
considered as valid only for galactic disc stars, at relatively high
metallicities.

\begin{acknowledgements}
We are grateful to the anonymous referee for a very useful, careful
and stimulating report, which greatly helped in the improvement of
this work. We acknowledge support from the Italian Ministry of
Research, under contract PRIN2006-022731, and from the Section of
Perugia of the National Institute for Nuclear Physics (INFN). This
study is also part of the preparatory work for the IRAIT Antarctic
telescope, of the Piano Nazionale delle Ricerche in Antartide
(PNRA). These studies on Antarctic Astronomy are profiting of the
financial support and cultural possibilities offered by the European
Coordinating Action ARENA (within the FP6 plan). \\
This research has made use of the SIMBAD database and the VizieR
service (CDS, Strasbourg, France), and the IRSA (NASA/IPAC
InfraRed Science Archive) database (USA), and the Astrogrid
database (UK). In particular archived data from the experiments
MSX, ISO-SWS and 2MASS were used. $\bullet$ The processing of the
science data of the Midcourse Space eXperiment (MSX) was funded by
the US Ballistic Missile Defense Organization with additional
support from NASA Office of Space Science. $\bullet$ The Infrared
Space Observatory (ISO) is an ESA project with instruments funded
by ESA Member States (especially the PI countries: France,
Germany, The Netherlands and UK). $\bullet$ 2MASS (Two Micron All
Sky Survey) is a joint project of the Univ. of Massachusetts and
the Infrared Processing and Analysis Center (IPAC) at California
Institute of Technology, funded by NASA and the NSF (USA).
\end{acknowledgements}

\begin{appendix}

\section{Period-Luminosity Relation \label{app1}}

\begin{figure*}[t]
\centering
{\includegraphics[width=7cm,angle=-90]{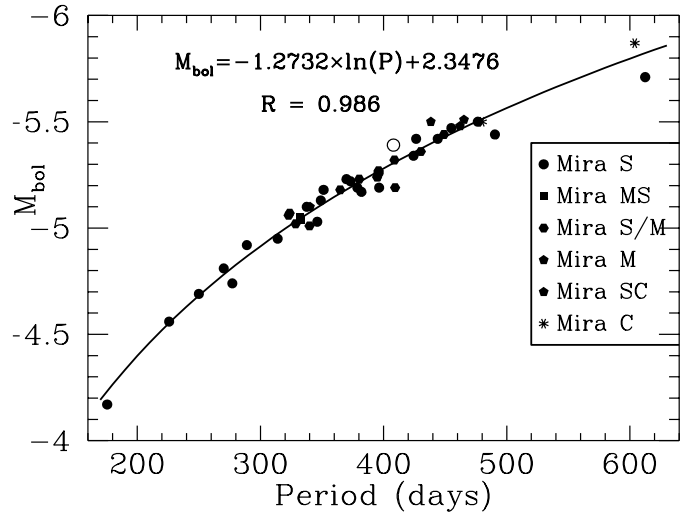}} \caption{The
average Period-Luminosity relation deduced for Miras (filled
symbols). The open circle refers to $\chi$ Cyg, whose magnitude,
deduced independently from its IR fluxes, its bolometric correction,
and its Hipparcos distance, fits well in the general relation. The
curve represents the written fitting relation, and R is its
regression coefficient. See text for further explanations.}
\label{fig9}
\end{figure*}

Mira stars have been alternatively suggested to be radially
pulsating either in the fundamental mode, or in the first
overtone \citep{tuch99,feast99}. For Semiregular variables
various overtones and the same fundamental mode are also
possible. The discussion on these properties has gone on for three
decades, although the large database made available by the MACHO
project \citep{alcock} seems to favor the fundamental mode for
Miras \citep{wood99}.

The non-unique pulsational properties of Semiregular variables
imply that there is not a unique $P-L$ relation for them
\citep{bedzij}. Instead, such relations for O-rich and C-rich
Miras have become increasingly accurate and statistically
relevant, to the point of forming now an invaluable tool in the
difficult task of determining the distances of Long Period
Variables. Many studies in the past were dedicated to correlate
$Log P$ with magnitudes obtained in the K filter
\citep{feast89,huwood,woodsebo}. This method, however, meets the
difficulty of the still remarkable light-curve amplitude at 2
$\mu$m. Relations directly connecting the period to the bolometric
magnitude \citep{feast89,whitelock94,whitelock06} are therefore
precious, although obviously more difficult to obtain.

Different relations are expected (and found) for C-rich and O-rich
objects \citep[see in particular][]{whitelock00,whitelock06}. For
our sample of S-stars, which have C/O $\leq$ 1, we can use the
$M_{bol} - P$ relation suggested for O-rich Miras by \citet{feast89}
and \citet{whitelock94}.

Another formula, established for Magellanic Cloud O-rich Miras, but
suitable also for the Galaxy in view of the universality of the
slope \citep{feast04}, was presented recently by
\citet{whitelock08}. This last relation is based on the K-magnitude.
It is difficult to establish criteria for choosing between the two,
so that we decided to use both, and then take an average of the
results. (We underline that both the original relations were derived
for O-rich sources and should therefore apply also to S stars, where
the ratio C/O does not reach unity). In particular, our procedure
was the following one. i) We firstly derived apparent bolometric
magnitudes from the bolometric corrections of Fig. \ref{fig3}, then
we applied the relations by \citet{whitelock94} to obtain an
estimate of the bolometric Magnitudes. Hence from the distance
modulus we could infer a first value for the distance. ii)
Separately, we applied the formula by \citet{whitelock08} to derive
the absolute K magnitude, then by comparison with the 2MASS
measurement we deduced a second estimate for the distance. The two
distance values were then averaged to get a final choice. This
choice for the stellar distance was then applied to our apparent
bolometric magnitudes, thus deducing our best estimate for the
absolute bolometric Magnitude.

Although certainly intricate, the above procedure should minimize
the systematic errors of each individual method.  The Magnitudes
thus deduced are plotted as a function of the period in
Fig.\ref{fig9}, thus providing our choice for the {\it average}
$P-L$ relation, descending primarily from the quoted works, but also
from our averaging technique and from our estimate of the bolometric
corrections. The derived Period-Magnitude relation can be expressed
by a fitting spline with a high accuracy, as shown on the same
graph. In the plot, we also included, as a comparison, a point (the
open circle) corresponding to the Mira star \emph{$\chi$ Cyg}. For
it we did not apply any $P - L$ relation: the magnitude was obtained
from directly scaling our apparent bolometric magnitude with the
known Hipparcos distance \citep{vleu2007}. Although it's only a
single point, its good accord with the trend defined by the other
sources gives us a consistency check for the technique.
\end{appendix}

\small

\onllongtab{11}{
\begin{longtable}{c c c c c c c}
\caption{\label{table:11} Sample D $-$ First part.}\\
\hline \hline
IRAS & Other & Stephenson & Coordinates & Spectral Type & Var. Type & Period \\
name & name & name & ICRS &   & (GCVS) & (GCVS) \\
\hline\hline
\endfirsthead
\caption{continued.}\\
\hline\hline
IRAS & Other & Stephenson & Coordinates & Spectral Type & Var. Type & Period \\
name & name & name & ICRS &   & (GCVS) & (GCVS) \\
\hline\hline
\endhead
\hline
\endfoot
04199+4950  &   \object{NX Per} &   CSS 91  &   04 23 39.35 +49 57 26.3 &   S   &   M   &   $-$ \\
06312$-$4351    &   \object{NU Pup} &   CSS 230 &   06 32 48.09 $-$43 54 13.6   &   S   &   M   &   $-$ \\
06468$-$1342    &   \object{DK CMa} &   CSS 264 &   06 49 06.38 $-$13 46 19.6   &   S:  &   M   &   $-$ \\
07571$-$2350    &   \object{EW Pup} &   CSS 440 &   07 59 16.04 $-$23 58 52.7   &   Se  &   M   &   $-$ \\
08348$-$3617    &   \object{WY Pyx} &   CSS 533 &   08 36 48.98 $-$36 27 36.5   &   S   &   M   &   $-$ \\
17186$-$2914    &   \object{V2434 Oph}  &   CSS 975 &   17 21 46.79 $-$29 17 34.0   &   S   &   M   &   $-$ \\
17589$-$1040    &   \object{V342 Ser}   &   CSS 1016    &   18 01 44.48 $-$10 40 25.2   &   S   &   M   &   $-$ \\
18375$-$1010    &   \object{V471 Sct}   &   CSS 1064    &   18 40 18.8 $-$10 07 29  &   S   &   M   &   $-$ \\
\end{longtable}
}
\onllongtab{12}{
\begin{longtable}{c c c c c c c c c c c c c}
\caption{\label{table:12} Sample D $-$ Second part.}\\
\hline \hline
Source & J & H & K & [8.8] & [9.8] & [11.7] & [12.5] & D & E &
Mid$-$IR Data Origin & Apparent bol. & I. $-$ E. \\
name & [Jy] & [Jy] & [Jy] & [Jy] & [Jy] & [Jy] & [Jy] & [Jy] &
[Jy] &   & magnitudes &   \\
\hline\hline
\endfirsthead
\caption{continued.}\\
\hline\hline
Source & J & H & K & [8.8] & [9.8] & [11.7] & [12.5]
& D & E &
Mid$-$IR Data Origin & Apparent bol. & I. $-$ E. \\
name & [Jy] & [Jy] & [Jy] & [Jy] & [Jy] & [Jy] & [Jy] & [Jy] &
[Jy] &   & magnitudes &   \\
\hline\hline
\endhead
\hline
\endfoot
NX Per  &   1.9 &   4.5 &   5.8 &   1.1 &   $-$ &   $-$ &   1.3 &   0.96    &   $-$ &   MSX &   7.61    &   $-$ \\
NU Pup  &   1.5 &   2.2 &   2.4 &   0.47    &   $-$ &   $-$ &   $-$ &   $-$ &   $-$ &   MSX &   8.58    &   I   \\
DK CMa  &   14.1    &   21.7    &   22.7    &   3.1 &   $-$ &   $-$ &   $-$ &   $-$ &   $-$ &   MSX &   6.06    &   I   \\
EW Pup  &   5.6 &   8.5 &   9.1 &   1.1 &   $-$ &   $-$ &   $-$ &   $-$ &   $-$ &   MSX &   7.04    &   $-$ \\
WY Pyx  &   34.8    &   58.0    &   62.8    &   13.6    &   $-$ &   $-$ &   8.8 &   6.4 &   3.4 &   MSX &   5.01    &   I   \\
V2434 Oph   &   21.2    &   53.1    &   65.9    &   11.4    &   $-$ &   $-$ &   10.3    &   6.9 &   4.0 &   MSX &   4.95    &   I   \\
V342 Ser    &   4.9 &   10.6    &   14.5    &   1.9 &   $-$ &   $-$ &   $-$ &   $-$ &   $-$ &   MSX &   6.55    &   I   \\
V471 Sct    &   8.6 &   25.6    &   30.8    &   8.1 &   $-$ &   $-$ &   9.6 &   6.8 &   4.1 &   MSX &   5.85    &   $-$ \\
\end{longtable}
}

\end{document}